\documentstyle[psfig]{mn}
\def\etal{{\rm et al.}}

\def\P3M{P$^3$M}
\def\vec#1{{\bf #1}}
\def\vecx{\vec{x}}
\def\vecn{\vec{n}}
\def\vecxi{\vec{x}_i}
\def\vecxj{\vec{x}_j}
\def\vecxjn{\vec{x}_{j\vec{n}}}
\def\jn{{j\vecn}}
\def\veck{\vec{k}}
\def\vecl{\vec{l}}
\def\vecX{\vec{x}^*}
\def\vecXjn{\vecX_{j\vec{n}^*}}
\def\refindent{\par \noindent \hang}
\def\paper#1#2#3#4#5{\refindent #1, #2, #3, #4, #5}

\def\and{, }
\font\mgn=cmti7
\def\eqnote#1{\marginpar{\mgn #1}}
\def\eqnote#1{{}}
 
\newfam\cyrfam
\font\tencyr=wncyr10 at 12pt 
\font\sevencyr=wncyr7 at 8.5pt \font\fivecyr=wncyr5 at 6pt
\def\cyr{\fam\cyrfam\tencyr}
\textfont\cyrfam=\tencyr \scriptfont\cyrfam=\sevencyr 
            \scriptscriptfont\cyrfam=\fivecyr
\def\comb{\mathop{\cyr Sh}}
\def\fracnum#1#2{\raise 2.1pt\hbox{$\scriptstyle #1$}\kern
-1.2pt/\kern -1.2pt \lower 2.1pt\hbox{$\scriptstyle#2$}\,}
\begin{document}

\title[Code paper] {Measuring the three-dimensional shear from
simulation data, with applications to weak gravitational lensing} 

\author[H. M. P. Couchman \etal] {H. M. P. Couchman$^1$\thanks{Email: 
couchman@coho.astro.uwo.ca}, Andrew J. Barber$^2$ and 
Peter A. Thomas$^2$ \\
{}$^1$Dept.~of Physics \& Astronomy, Univ.~of Western Ontario,
London, Ontario, N6A 3K7, Canada\\ 
{}$^2$Astronomy Centre,
University of Sussex, Falmer, Brighton, BN1 9QJ
}

\date{Accepted 1998 ---. Received 1998 ---; in original form 1998 ---}

\maketitle

\begin{abstract}

We have developed a new three-dimensional algorithm, based on the
standard P$^3$M method, for computing deflections due to weak
gravitational lensing. We compare the results of this method with
those of the two-dimensional planar approach, and rigorously outline
the conditions under which the two approaches are equivalent. Our new
algorithm uses a Fast Fourier Transform convolution method for speed,
and has a variable softening feature to provide a realistic
interpretation of the large-scale structure in a simulation. The
output values of the code are compared with those from the Ewald
summation method, which we describe and develop in detail. With an
optimal choice of the high frequency filtering in the Fourier
convolution, the maximum errors, when using only a single particle,
are about 7 per cent, with an rms error less than 2 per cent. For
ensembles of particles, used in typical $N$-body simulations, the rms
errors are typically 0.3 per cent. We describe how the output from the
algorithm can be used to generate distributions of magnification,
source ellipticity, shear and convergence for large-scale structure.

\end{abstract}

\begin{keywords}
Galaxies: clustering --- Cosmology: miscellaneous --- Cosmology:
gravitational lensing --- Methods: numerical --- Large-scale structure
of Universe
\end{keywords}

\section{INTRODUCTION}
 
Procedures for the generation of cosmological $N$-body simulations
have become increasingly sophisticated in recent years. We have now
developed an algorithm for assessing in great detail, and with
considerable speed and accuracy, the effects of the large-scale mass
distributions within these simulations on the passage of light from
sources at great distances.

\subsection{Previous work}

There are numerous methods for studying such
`weak' gravitational lensing, the most common being `ray-tracing,' in
which individual light rays are traced backwards from the observer,
and the deflections occuring at each lens-plane are calculated. The
lens-planes are two-dimensional projections of the mass content within
a small redshift interval, usually equal to the simulation box depth.

Schneider and Weiss (1988b) have used this method by shooting,
typically, $10^8$ rays through the lens-planes to strike the source
plane within a chosen square region, called the detection field. Each
plane is divided into $51^2$ pixels, typically, and the particles
(stars in this case) are categorised as near or far for computational
purposes. The deflection by stars further away than about 8 pixel
dimensions from the centre of each pixel is approximated in a Taylor
series, whilst the deflections due to the nearby stars are computed
individually. The rays are shot through a cylinder with a radius such
that the rays meet the source plane in or near the detection
field. They claim that the resulting amplification factors are hardly
affected by edge effects caused by strong lensing of rays outside the
shooting cylinder, but which might strike the detection field. The
amplification factors are determined directly from the mapping of the
rays onto the pixels of the detection field.

Jaroszy\'{n}ski et al. (1990) evaluate the matter column density in a
matrix of pixels for each of the lens-planes, based on their
simulation boxes. By making use of the periodicity in the particle
distribution orthogonal to the line of sight, they are able to arrange
for each ray traced to be centralised within planes of one full period
in extent. In this way, the deflections on each ray take account of
all the matter within one complete period. They calculate, not
deflection angles, but the two two-dimensional components of the
shear, (see Section 5.3), as ratios of the mean convergence of the
beam. To do this, they assume that the matter in each of the $128^3$
pixels resides at the centre point of each pixel. To follow the
shearing across subsequent planes they recursively generate the
developing Jacobian matrix for each ray, in accordance with the
multiple lens-plane theory, (see Section 5.3).

Wambsganss (1990) uses the `ray-tracing' method to study microlensing,
Wambsganss, Cen and Ostriker (1998) use it with cosmological $N$-body
simulations, and Wambsganss et al. (1997) use the
method for studying the dispersion in type Ia supernov\ae. They
randomly orient each simulation box, and project the matter
contained within each onto a plane divided into pixels. They choose
the central $8h^{-1}$Mpc $\times~8h^{-1}$Mpc region through which to
shoot rays, ($h$ is the Hubble constant in units of
100~km~s$^{-1}$~Mpc$^{-1}$), but account for the deflections of the rays
in terms of all the matter in the plane of $80h^{-1}$Mpc $\times
~80h^{-1}$Mpc. However, to speed up the computation, a hierarchical
tree code in two dimensions is used to collect together those lenses
(pixels) far away, whilst treating nearby lenses
individually. The matter in each pixel, which measures $10h^{-1}$kpc
$\times~10h^{-1}$kpc, is assumed to be uniformly spread. The multiple
lens-plane theory, (see Section 5.3), is used for large numbers of
rays to compute the mappings of images and sources, the distribution
of magnifications, and statistics of angular separations of multiple
images.

Marri and Ferrara (1998) select a total of 50 planes, evenly spaced in
redshift space, between redshifts of $z=0$ and $z=10$. Their
two-dimensional matter distribution on each plane consists of point
masses without softening, so that their approach produces very high
magnifications (greater than 20) for each of their chosen cosmologies,
using the `ray-tracing' method.

Premadi, Martel and Matzner (1998) have used 5 different sets of
initial conditions for each $N$-body cosmological simulation, so that
the plane projections of each simulation box can be chosen randomly
from any one of the 5, and then randomly translated, using the
periodic properties of each box. In this way they are able to avoid
correlations in the large-scale structure between adjacent boxes. A
considerable improvement to the `ray-tracing' method has then been
made. They solve the two-dimensional Poisson equation
on a grid, and invert the equation using a two-dimensional Fast
Fourier Transform (FFT) method, to obtain the first and second
derivatives of the gravitational potential on each plane. From this
data, cumulative deflections and the developing Jacobian matrix can be
obtained, which provides the data for determining overall
magnifications.

An alternative to the conventional form of `ray-tracing' was
introduced by Refsdal (1970), who used the `ray-bundle' method. The
principle here is to trace the passage of a discrete bundle of light
rays as it passes through the deflection planes. The advantage of this
method is that it provides a direct comparison between the shape and
size of the bundle at the observer and at the source plane, so that
the magnification, ellipticity and rotation can be determined
straightforwardly.

Fluke, Webster and Mortlock (1998a, b) use the `ray-bundle' method,
and have found that bundles of 8 rays plus a central ray adequately
represent a circular image which is projected backwards. They shoot
large numbers of bundles through two-dimensional planes, as above,
formed by projecting their simulation cubes, which have comoving sides
of $60h^{-1}$Mpc. The shooting area is limited to $50\arcsec \times
50\arcsec$, and the deflection angles are calculated by considering
the matter contained within a chosen radius, typically $15h^{-1}$Mpc
from the central ray.

A novel approach to weak gravitational lensing has been used by Holz
and Wald (1997). They lay down a set of spheres between the observer
and source, each containing an individual probability distribution of
matter, in which a `Newtonianly perturbed' Robertson-Walker metric is
used.  A scalar potential, related to the density perturbations, can
then be evaluated, and this allows integration along straight lines
through each sphere, to determine the angular deviations and shear.

Tomita (1998a) evaluates the potential at some 3000 positions between
the observer and a source at $z=5$, by using the periodic properties
of each simulation cube to position it such that each evaluation
position is centrally placed in the appropriate cube. To trace the
paths of the light rays, they solve the null-geodesic equations, and
use an analytical expression to determine the average potential
through the interval between each pair of evaluation positions.

\subsection{Motivation}
 
Our motivation for the development of an algorithm to apply in
three-dimensions has stemmed from concern with possible
limitations in the two-dimensional planar approaches to weak
gravitational lensing. We therefore considered the following.
 
First, we wanted to investigate rigorously the conditions for
equivalence of results obtained from three-dimensional realisations
and two-dimensional planar projections. We show in Appendix B, that
the shear values\footnote{Note that, throughout this paper, we refer
to the elements of the matrix of second derivatives of the
gravitational potential as the `shear' components, although, strictly,
the term `shear' refers to combinations of these elements which give
rise to anisotropy.} at a point in the two-dimensional projection are
equal to the integrated three-dimensional values along the line of
sight through one period (or cube dimension), in general, only if the
distribution of mass is periodic along the line of sight, and the
angular diameter distances through the depth of one period are assumed
to be constant.
 
Second, we wanted a method which would unambiguously provide
accurate values for the shear components, as if the contribution from
all matter, effectively stretching to infinity, was included. Errors
may occur in other methods if the contribution from matter within only
a finite radius of the evaluation position is counted. For example,
whilst Jaroszy\'{n}ski at al. (1990) include the matter contained
within a plane of one complete period, Wambsganss (1990), Wambsganss
et al. (1997) and Wambsganss et al. (1998) introduce a slight bias,
because rays near the edge of their shooting area are closer to the
edge of the single period plane than rays near the centre of the
shooting area. Fluke et al. (1998a, b) typically consider a region
of radius $15h^{-1}$Mpc in simulations with a period of
$60h^{-1}$Mpc. We investigate how quickly the shear component values
converge to their true values, as increasing volumes of matter are
included surrounding the evaluation position, and we report our
findings in Section 4.

Third, to achieve shear values consistent with those in a realistic
universe, it is necessary to deal with the `peculiar potential,'
$\phi$, which is related to the full gravitational potential, $\Phi$,
through the subtraction of a term depending upon the mean
density. This approach, which we describe fully in Appendix A, ensures
that we deal only with light ray deflections arising from departures
from homogeneity; in a pure Robertson-Walker metric we would want no
deflections. The approach is equivalent to requiring that the net
total mass in the system be set to zero, so that for every mass there
is a balancing uniform negative background mass. This net zero mass
requirement is achieved very simply in systems of particles which are
periodically distributed in all dimensions, and we are therefore able
to accommodate it easily in our algorithm.

Fourth, we wanted to be able to apply our algorithm to cosmological
simulations which were representative of the large-scale structure in
the universe, rather than distributions of point masses. This has
frequently been attempted by assuming each particle to be softened
with the profile of an isothermal sphere. We wanted to improve on this
by allowing the softening for each particle to reflect the environment
in which it is located, so that large clusters and other dense
structures dominate the light deflections, whilst the effects of
isolated particles are minimised. This motivated us to consider the
introduction of a variable softening facility to our algorithm.

Finally, a three-dimensional approach allows the use of the
appropriate angular diameter distances at every single evaluation
position within the three-dimensional realisation. The two-dimensional
methods discussed above necessarily assume that all the lensing
mass within a given plane is at the same angular diameter distance,
because the overall depth of the lens is considered to be small
compared with the observer-lens, observer-source, and source-lens
angular diameter distances. However, the depth of a single simulation
cube employed in weak lensing, (in our case $100h^{-1}$ Mpc), is not
insignificant. By using the `thin plane approximation,' in which
matter in a small redshift interval is projected onto a plane, we are
able to show that errors may be introduced. It is not possible to
quantify these errors in general, because they will vary from
simulation to simulation, depending on the specific mass
distributions. However, we show in Section 4.2 that the scaling
factors for the computed shear components can vary by as much as 9\%
through the depth of those simulation boxes contributing the most to
the magnification.
 
The considerations above finally led us to develop an efficient Fast
Fourier Transform (FFT) programme, whose data output could be
manipulated with the appropriate angular diameter distances at every
evaluation point. The primary output of the programme is the matrix of
second derivatives of the gravitational potential at each of the
specified evaluation positions within a periodic three-dimensional
distribution of smoothed particle masses.

\subsection{Outline of paper}
 
In section 2, we describe the general principles employed in the
standard Particle-Particle, Particle-Mesh (P$^3$M) algorithm, and then
how it has been extended for the evaluation of the shear
components. We explain the introduction of variable softening into the
code, which allows each particle to be represented as a distributed
mass. This variable softening smooths away the high frequency
information in very high density clumps, thus avoiding strong
scattering, and also allows particles in low density regions to be
spread more widely to give more realistic density values.
 
In section 3, we describe our testing procedures for the code. The
first of these compares the computed shear components at a large
number of points surrounding a single massive particle, with values
derived from the Ewald (1921) summation technique. The second test
compares values of the normalised trace of the shear matrix with
density values derived using an independent method, a smoothed
particle hydrodynamics, SPH, algorithm.

Section 4 emphasises two advantages of our new algorithm. First, we
demonstrate the slow convergence of shear values to their true values,
as increasingly large volumes of matter are included surrounding an
evaluation position. This suggests the need, in general, to include
the effects of matter well beyond a single period transverse to the
line of sight. Secondly, we show that, by considering all the matter
in a cubic simulation to be at the same angular diameter distance,
sizeable errors may be introduced to the absolute shear values, and to
calculations of the magnification along a line of sight.
 
Section 5 describes the cosmological $N$-body simulations we are using
for the application of the new algorithm, and explains our choice of
the appropriate minimum value for the variable softening.  The
multiple lens-plane theory (described by Schneider, Ehlers \& Falco,
1992) is summarised for the determination of magnification
distributions along large numbers of lines of sight through the
simulations. We then provide some preliminary results to show the
efficacy of the method. These include (a) plots of the magnification
as it develops along a line of sight, (b) values for the shear and
convergence in a given cosmological simulation, and (c) distributions
of the magnification due to weak lensing for isolated simulation
boxes. Finally, we outline our proposed future work, which will link
simulations together to provide a complete realisation from a distant
source to an observer in the present epoch. This will enable us to
compare the results from different cosmologies.
 
Section 6 summarises our conclusions about the algorithm and its
applicability.

In Appendix A, we describe how the peculiar potential relates to that
in a universe with large-scale homogeneity. We show how use of the
peculiar potential, which takes account of departures from
homogeneity through the subtraction of a term including the mean
density, allows the shear to be correctly computed.
 
In Appendix B, we investigate rigorously the equivalence between
two-dimensional and three-dimensional periodic approaches, in the
absence of discrete angular diameter distances within the
realisations. The treatment details the conditions under which the
two approaches may be considered to be equivalent.
 
In Appendix C, we summarise the Ewald (1921) summation method, which
we have used to assess the accuracy of our new code. We describe the
method in outline, and compare the treatment with the P$^3$M
method. Finally, we develop the equations for the summation method
which we have used in the testing of the results from our new
algorithm.

\section{IMPLEMENTATION}

In this section we describe the numerical method used for measuring
the local 3-dimensional shear in simulation data. The technique is an
extension of the standard \P3M\ algorithm familiar from cosmological
particle codes. We begin with a brief review of the \P3M\ method and
then describe how it has been extended for the shear calculation.

\subsection{The \P3M\ algorithm}

The \P3M\ algorithm was developed in the context of particle
simulations of plasmas by Hockney, Eastwood and co-workers, (see
Hockney \& Eastwood, 1988, for a full description), as an efficient
method for calculating the pairwise interactions of a large number of
particles. In the cosmological context, with the method being used to
calculate forces arising from a large number of self gravitating
particles, the method has two important attractions. First, for a
nearly uniform distribution of particles, the computational cost is of
order $N{\rm log}_2 N$ where $N$ is the number of particles, rather
than the normal $O(N^2)$ scaling behaviour expected for a na\"{\i}ve
computation of the forces on $N$ particles from each of their $(N-1)$
neighbours. The second attractive feature for cosmology is that the
method, in its standard form, has periodic boundary conditions, and
thus lends itself naturally to the simulation of a small part of the
universe with the remainder being approximated by periodic images of
the fundamental volume.

The key idea in the method is to decompose the pairwise interparticle
force into a long-range and a short-range component which together sum
to the required force. With a suitable choice of the decomposition we
can ensure that the short-range force is compact (that is, it is
non-zero only within a finite radius, i.e., the `search radius') and
that the long-range component has a band-limited harmonic content such
that it can be accurately represented by sampling with a regular grid
of a convenient mesh size. The total force is then accumulated on
particles by summing directly a contribution corresponding to the
short-range component of the force from nearby particles within the
search radius, together with the long-range component which is
interpolated from a smoothly varying force derived from the regular
mesh.

In practice, the calculation of the short-range force (from the direct
particle-particle (PP) sum over near neighbours) and the long-range
force (from the particle-mesh (PM) field calculation) depend on each
other only to the extent that the accumulated force from the two
should sum to the required total force. The two calculations may be
performed in either order: here we shall describe the PM calculation
first.

The heart of the PM calculation is the solution of Poisson's equation
on a grid via a rapid elliptic solver; the method used here is Fast
Fourier Transform (FFT) convolution. A mesh-sampled density is
obtained by smoothing the particle distribution onto a regular grid
with an appropriate kernel. The properties of the kernel are chosen to
filter the high frequencies present in the distribution so that the
smoothed distribution may be adequately sampled by the mesh. The mesh
potential is then obtained from the mesh density by FFT
convolution. An advantage of using an FFT method is that it is
possible to substantially reduce translational and directional errors
in the mesh-computed quantities by judicious adjustment of the Fourier
components of the Green's function. For the standard force
calculation, the components of the Green's function are optimized such
that the rms deviation of the computed force from the desired force is
minimized. Full details of these techniques may be found in Hockney \&
Eastwood~(1988). A key feature of the method is that the
Fourier-transformed density field may be smoothed by using a high
frequency filter. This suppresses aliasing and leads to a more
accurate pairwise force; that is one which has less positional and
rotational dependence relative to the grid.

Derivatives of the potential may then be obtained at mesh-points using
standard finite difference techniques. A 10 point differencing
operator is used here to minimize directional errors in the computed
differences, (see Couchman, Thomas \& Pearce, 1995). Values of the
potential and its derivatives at arbitrary points in the computational
domain are then obtained by interpolation from the mesh values. (Using
the same kernel for interpolation as was used for particle smoothing
ensures that particles do not experience self-forces in the standard
\P3M\ method.)

The accumulation of the PP component of the force on a particle from
near neighbours is achieved by re-gridding the particles onto a mesh
which has a cell size such that the side is equal to the radial
distance at which the short-range force falls to zero. This mechanism
enables the neighbours contributing to the short-range force to be
found efficiently by searching over the cell in which the particle in
question lies and its 26 neighbouring cells. A disadvantage of this
technique is that as particle clustering develops in a simulation the
average number of neighbours rises, causing the method to slow as the
number of PP contributions which must be computed increases. A
technique to overcome this deficiency in simulation codes has been
developed, (Couchman, 1991), but the problem will not be of concern in
this paper where we shall be concerned only with limited clustering.

The method leads to accurate interparticle forces with the force error
(arising from the mesh aliasing) being controlled by the degree of
high frequency filtering employed in the Fourier convolution. A
greater degree of attenuation of the high frequency components reduces
the error but leads to a `smoother' mesh force. This requires that the
direct-sum search be performed out to larger radii, which in turn
requires a search over a greater number of particles leading to a
slow-down in the execution of the code.

The method may be described schematically in the following
terms. Suppose that the total pairwise potential required is $\varphi
= \varphi(r)$, where $r$ is the radial distance from a particle. Then
we compute this as $\varphi=\varphi_{\rm PP} +
\varphi_{\rm PM}$, where $\varphi_{\rm PP}(r) = 0$ for $r>r_c$, is the
PP component and $\varphi_{\rm PM}$ is the mesh part. The functional
form of $\varphi$ is Coulombic on large scales (neglecting for the
moment the effect of the periodic images), with perhaps a softening at
small scales to allow for the fact that each particle may
represent a very large astrophysical mass, and to ameliorate certain
numerical problems in the simulation code such as two-body
scattering. 

\subsection{Adaptation of the \P3M\ algorithm for the calculation of
the shear components}

The \P3M\ method computes forces, or first derivatives of the
potential, $\phi$, at a point by splitting the contribution of the
density distribution into two components as described above. The
potential itself is also computed as a simulation diagnostic in many
standard \P3M\ implementations using the same splitting technique. In
principle, any other non-local function computed from the field may be
treated in the same way, and this is the approach taken here for the
shear components, $\partial^2\phi/\partial x_i \partial x_j$. The
implementation details specific to the calculation of the shear values
will be discussed here.

The short-range part of the shear field at a point is accumulated
directly from neighbouring particles from the appropriate Cartesian
projections of the analytic function:
\begin{equation}
\left({\partial^2\phi\over\partial x_i \partial x_j}\right)_{\rm
PP}= \sum {\varphi_{\rm PP}^\prime\over x}\delta_{ij} +
\left(\varphi_{\rm PP}^{\prime\prime} - {\varphi_{\rm PP}^\prime\over
x}\right) {x_i x_j\over x^2},
\end{equation} 
where the sum is over all
neighbour particles and $x$ is the separation of a neighbour particle
from the point at which the shear is desired. (We have used the prime
notation to denote derivatives with respect to radial separation.) The
short-range part of the field may be computed to machine accuracy.

The long-range part of the shear field is derived by taking a second
difference of the force values as computed in the standard \P3M\
method. The only difficulty is that differencing the mesh field
magnifies the noise which is present as a result of aliasing. Reducing
this noise requires more filtering in the Fourier domain with a
corresponding increase in the short-range cutoff, $r_c$. 

Optimization of the Green's function appropriate for the shear
calculation is done in a manner similar to that used for the force
calculation. We minimized the sum of the squares of the deviations of
all nine components of the shear, although a number of other reasonable
possibilities exist. Minimizing the deviation of the diagonal
components, for example, produced results that were little different.

It would be possible to compute the mesh shear components by inverse
Fourier transform of $-k_i k_j \tilde\phi({\bf k})$ directly for each
$i$, $j$ such that $1\le i\le j\le3$, thus avoiding the real-space
differencing. It would still be necessary to filter the field,
however, and for the differencing operator used, the very small
increase in accuracy would not justify the added computational cost of
several further FFTs.

\subsection{Particle softening}

An important feature of numerical particle codes is the use of
particle `softening'. The effect of this is that each particle in
the code represents not a point mass but a distributed mass with some
given (fixed) radial profile. Softening is introduced primarily to
avoid artificial (or numerical) relaxation; that is, close two-body
encounters leading to spurious energy redistribution in the system.
Since in most simulations we are attempting to model the cosmic matter
density as a collisionless fluid, this is a useful approach. (Note that
the particle softening referred to here is distinct from the high
frequency filtering, or smoothing, employed in the PM part of the
calculation.)

In numerical particle simulation codes it is usual to employ a global
softening for all particles (which may, however, vary in time). As the
particle distribution evolves and particles cluster the low density
regions are represented by fewer particles. In a simulation computing
interparticle forces to evolve the distribution of particles, this is
of little consequence. However, if we wish to compute the shear values
at some position on a ray passing through a low density region it may
never come within the softening of the widely spread particles in the
region. Since we are interested in the trace of the shear matrix,
which is equivalent to the density, this would suggest that the
density at this point was zero. (In fact the density would be negative
since the total mass in the periodic simulation cube must be zero.)
This is unrealistic and does not accurately represent the matter
density in these regions. Increasing the softening would ameliorate
this situation in the voids but would smooth away the high frequency
information present in regions where the particles have clustered into
high density lumps.

The approach we have taken is to employ a variable softening, such
that a particle in a region of low particle density has a `size' which
is greater than that of a particle in a high density region. We have
chosen a criterion similar to that used in hydrodynamic simulations
using the SPH method (e.g., Gingold \& Monaghan, 1977). Each particle
is chosen to have a softening such that its sphere of influence is
proportional to the distance to its 32nd nearest neighbour. Given
these values the code appropriately modifies the short-range (PP)
calculation to increment the shear components at a given point taking
into account the varying sizes of the neighbouring particles.

\section{PERFORMANCE AND TECHNICAL ASSESSMENT}

\subsection{Pairwise shear tests}

A minimal check of the technique may be made by computing the shear
components at a large number of points surrounding a single massive
particle. This is a useful test because the result is known
analytically and it provides an immediate assessment of the errors
present in the method.

The test was made as follows. A single massive particle was placed at
a random location in a mesh cell and the code used to measure the
shear components at 16384 surrounding points located randomly in
direction and distributed in radius such that there was an equal
number of points per equal logarithmic increment in radius. The test
was then repeated a further 34 times using the same evaluation
positions, but different locations for the test particle designed to
adequately sample the mesh cell. The shear components measured at each
location were then compared with the true values derived from the
Ewald (1921) summation technique as described in Appendix C.  These
comparisons are plotted in Figure 1.

\begin{figure}
$$\vbox{
\psfig{figure=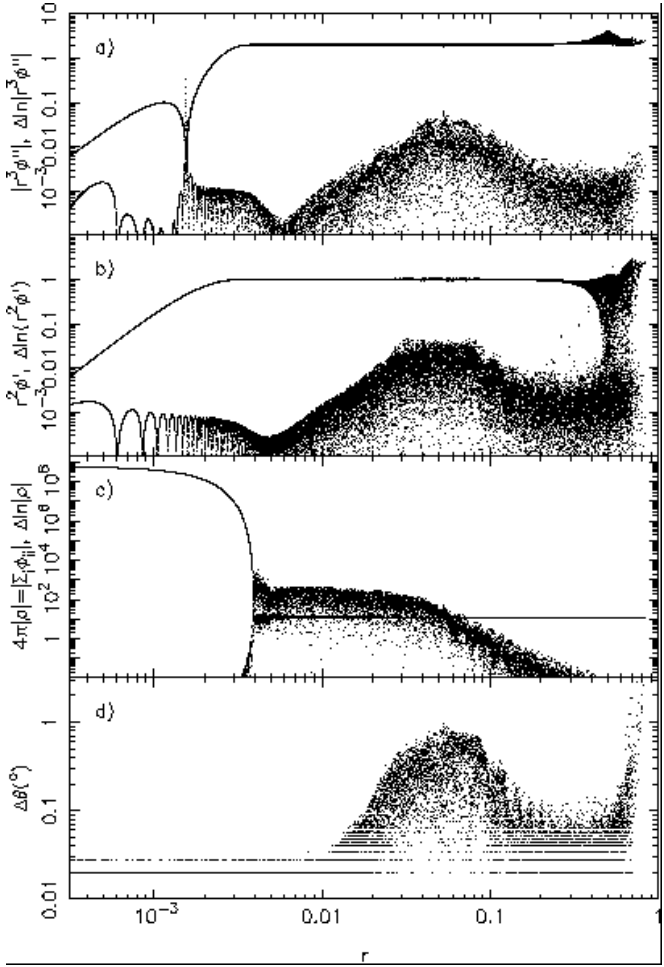,width=8.7truecm}
}$$
\caption{Computed shear components and errors in these components
arising from the single particle test described in Section 3.1. For a
complete description of the test and of the features shown in the
figure refer to the text. All separations, $r$, are expressed in units of
the system's period.}
\label{fig:hughserrors}
\end{figure}

Panel a) of Figure 1 shows the absolute value of the radial component
of the shear (solid line) as well as the fractional error in this
quantity (scattered dots). Panel b) shows the same quantities for the
two transverse components. In both cases the measured values have been
multiplied by an appropriate power of the separation, $r$, such that a
pure Coulombic potential would show no radial variation.  For the
filtering chosen, the maximum error in these quantities is
approximately 7 per cent, and occurs near the mesh scale used. The
test was performed using a mesh of $64^3$ cells, but the errors are
essentially independent of the mesh size for typical values of the
softening. The cyclic error at small separations arises because linear
interpolation into a look-up table is used for the short-range
force. The few points lying above the main scatter of errors in panel
b) are due to one of the transverse components becoming very small
with the consequence that the fractional error can become large. The
rms error in our test on a single particle is less than 2 per cent.

Panel c) plots the absolute value of the trace of the Ewald-computed
shear matrix and the fractional error of the computed trace
values. The particle softening chosen for this test was 0.002 in units
of the box side dimension and, because of the definition of the particle
softening employed, this corresponds to the particle having a radial
extent of 0.004. The true value of the trace for smaller separations
thus reflects $4\pi$ times the density of the particle as required by
Poisson's equation. At larger separations the true density is -1
(since the total mass in the system has to be zero),
and thus the Ewald-computed shear is $4\pi$. Large errors occur in
the trace values simply because the shear values individually are many
orders of magnitude larger than $4\pi$ and each is in error by a few
percent; we thus cannot expect the sum of these large values to cancel
to high precision to give the required result. This fact was
one of the primary motivations for using a variable softening in which
every position at which the shear was measured would lie within the
effective radius of some particle. In this case we may expect that the
computed values of the trace will have roughly the same fractional
error as shown in panels a) and b). This is discussed and tested
further in the next section.

Finally, panel d) shows the directional error between the principal
eigenvector of the computed shear matrix and the separation
vector. (The banding for small directional errors is due to numerical
quantization in the computation of the directional error.) It is
apparent that directional errors are very small in this method.

Note that the errors in the shear values computed from an {\it
ensemble} of particles are, in general, much smaller than the pairwise
errors shown in Figure~1. (For ensembles used in typical $N$-body
simulations, the rms force errors are typically 0.3 per cent.) This is
because the Fourier representation of a general particle distribution
will have a smaller high frequency content than the equivalent
representation of a single massive particle and can thus be
represented better by a given fixed grid. Only if the PP and PM forces
almost exactly cancel will the {\it fractional} errors be larger
although in this case the absolute error will be small. There will be
no appreciable change in the error values with larger mesh
configurations normally used with $N$-body simulation data.

\subsection{Comparison of measured trace and overdensity for a
distribution of particles}

For a more realistic test of the code, we compare the computed shear
trace with the density evaluated using a standard SPH algorithm for
one of our cosmological simulation boxes, described in Section
5.1. The SPH programme evaluates a parameter, $l$, at each particle in
the simulation box, representing half the distance to the 32nd nearest
neighbour. This parameter defines the volume for the density
calculation, and the same parameter is applied in
the shear code to establish an appropriate value of the softening for
the particle. In addition, a specific smoothing function may be used
to distribute the mass throughout the volume so defined.

To make a suitable comparison with the shear trace values, we have
computed overdensity values from these densities, and compared the
ratio of the overdensity and the trace with the overdensity values. In
Figure 2, we plot the average value of this ratio in each overdensity
bin. We have used a minimum value of 0.0005 for the variable
softening, and analysed the data from 10000 particle positions.

\begin{figure}
$$\vbox{
\psfig{figure=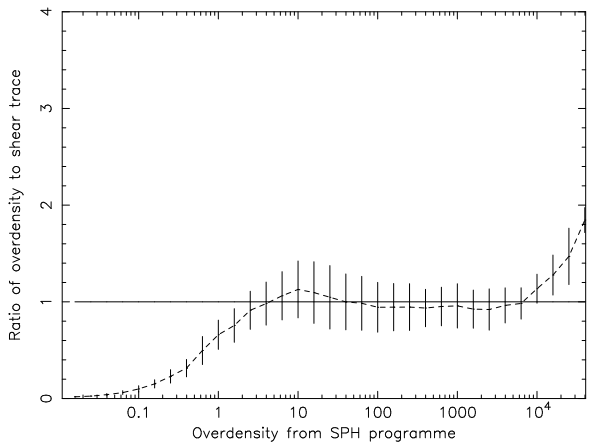,width=8.7truecm,angle=270}
}$$
\caption{Ratio of overdensity from an SPH programme to the normalised
shear trace, against the binned overdensity. The
average values of the ratio are shown by the broken line; the full
line is drawn at the value of unity for comparison.}
\label{fig:sigmabar_20.ps}
\end{figure}

Because of the very different ways in which the densities are
determined (from the shear trace in the new code, and from particle
numbers in the SPH programme, and the differing shapes of the
softening functions used), we expect some dispersion in the values at
all densities, and this is indicated by the 1$\sigma$ error bars. The
form of the plot is easily understood throughout its entire range.  At
low densities, the SPH density values are underestimated because
isolated particles do not have the requisite number of nearest
neighbours within the particle mesh.  At high densities, the particle
softenings will be at the same minimum level for all the particles,
retarding the amount of increase in the shear trace values as the real
density continues to rise. This effect causes the upturn shown in
Figure 2, which occurs at the overdensity value of $4.5 \times 10^3$
for a minimum softening of 0.0005.

The equality of the particle overdensities and shear trace values over
more than three decades in density, gives us considerable confidence in
the use of the shear algorithm generally, and for trace values
determined at particle positions, or within the softening range.

\section{SOME ADVANTAGES OF THE THREE-DIMENSIONAL METHOD}

\subsection{Convergence to limiting values}
 
In the Introduction we outlined some of the considerations we made
before developing the new three-dimensional algorithm. We expressed
concern that shear values in general may converge only slowly to their
true limiting values as increasingly large volumes of matter are
included around an evaluation position. If so, it would be essential
either to take full account of the periodicity of matter orthogonal to
the line of sight, or to include the effects of matter over
considerable distances from the line of sight. 

We investigated this rate of convergence of the shear values for one
of our simulation boxes, (which we describe in Section 5.1). By using
a straightforward direct-summation method for the particle
contributions to the shear, we evaluated one of the off-diagonal
two-dimensional shear components as we progressively added mass out to
a radial extent of 2.5 box units. Beyond a radius of 0.5 from the
central position, the particles were laid down with the periodicity of
the fundamental volume. The depth was one box unit throughout, because
of the inbuilt periodicity along the line of sight. (We show in
Appendix B that provided there is periodicity along the line of sight,
the two-dimensional shear values will equate with the
three-dimensional results integrated over a single period.)

Figure 3 clearly shows that by including the matter within a single
period only, (to a radius of 0.5), values for the shear components
will, in general, be seriously in error. Of course, different
simulations and particle distributions will display different rates of
convergence to the limiting values. However, it is quite clear that by
making correct use of the periodicity in simulations (as an
approximation to the distribution of matter outside of each simulation
cube), together with the net zero mass requirement, more realistic
component values are achieved. Other approaches which do not employ
these two conditions may suffer from inadequate convergence to the
limiting values.
 
\begin{figure}
$$\vbox{
\psfig{figure=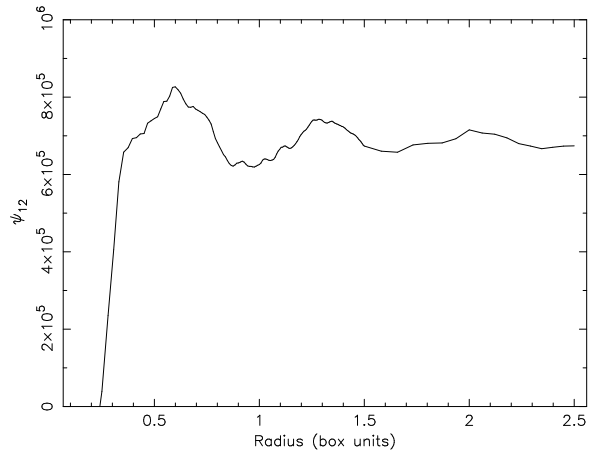,width=8.7truecm,angle=270}
}$$
\caption{One of the shear components, $\partial^{2} \phi /\partial
x\partial y$, as a function of the radial extent of matter included.}
\label{fig:twod9_25_031.ps}
\end{figure}

\subsection{The effects of angular diameter distances}
 
Our three-dimensional approach allows the use of the appropriate
angular diameter distances at every single evaluation position. This
is not possible in two-dimensional approaches, where it is assumed
that all the lensing mass is projected onto a plane at a single
angular diameter distance.

By definition, the angular diameter distance of a source is the
distance inferred from its angular size, assuming Euclidean
geometry. In an expanding universe, therefore, the angular diameter
distance becomes a function of the redshift of the source (and
of the observer). In addition, the inclusion of excess matter within
the beam causes the beam to become more focussed, and makes the source
appear closer than it really is. By considering the universe to be
populated by randomly distributed matter inhomogeneities, but
resembling the Robertson-Walker, Friedmann-Lema\^{i}tre model on large
scales, (see Schneider, Ehlers and Falco, 1992), a second order
differential equation is obtained for the angular diameter distance,
$D$, in terms of the density parameter, $\Omega$, for the universe, and
the redshift, $z$, of the source:
\begin{equation}
\left(z+1 \right) \left(\Omega z+1\right)
\frac{d^2D}{dz^2}+\left(\frac{7}{2}\Omega z + \frac{\Omega}{2} + 3
\right) + \frac{3}{2}\Omega D = 0.
\end{equation}

Dyer and Roeder (1973) made assumptions about the type of matter
distribution to obtain a more general and practical equation. They
assumed that a mass fraction, $\bar{\alpha}$, (called the smoothness
parameter), of matter in the universe is smoothly distributed, and
that the fraction $(1-\bar{\alpha})$ is bound into clumps. Then the
equation for the angular diameter distance becomes
\begin{eqnarray}
\lefteqn{\left(z+1 \right) \left(\Omega z+1\right)
\frac{d^2D}{dz^2}+\left(\frac{7}{2}\Omega z + \frac{\Omega}{2} + 3
\right)} \nonumber \\
 & & \hskip 1.0 in + \left(\frac{3}{2}\bar{\alpha}\Omega + 
\frac{\mid \sigma \mid^2}{(1+z)^5} \right) D = 0,
\end{eqnarray}
in which shear, $\sigma$, is introduced by the 
matter distribution around the beam. They considered 
the following scenarios for the application of this equation. First,
they considered a universe in which all the matter is bound into
clumps, so that $\bar{\alpha}=0$, and in which the light beam passes
far away from the clumps. This is described as light propagating
through an `empty cone,' and gives rise to maximal divergence of the
beam. The second scenario is more general and practical, in that it
uses an intermediate value for the smoothness parameter
($0<\bar{\alpha}<1$), but still requires the beam to pass far away
from the clumps. In this case the beam contains a proportion of the
smoothed matter distribution which introduces convergence, and hence
some degree of focussing. The third scenario has $\bar{\alpha}=1$,
i.e., an entirely smooth universe. Here the smooth matter distribution
is present within the beam, giving a `full cone,' or `filled beam'
approximation.

In all of these scenarios the term including the shear in equation (2)
is minimised, so that the final `Dyer-Roeder equation' becomes
\begin{equation}
\left(z+1 \right) \left(\Omega z+1\right)
\frac{d^2D}{dz^2}+\left(\frac{7}{2}\Omega z + \frac{\Omega}{2} + 3
\right) + \frac{3}{2}\bar{\alpha}\Omega D = 0,
\end{equation}
and can be solved for different values of $\Omega$
and $\bar{\alpha}$. For $\Omega =1$ and $\bar{\alpha}=1$ (filled
beam), Schneider et al. (1992) quotes the result for the angular
diameter distance between an observer at redshift $z_1$, and a source
at redshift $z_2$, as
\begin{equation}
D(z_1,z_2) =
\frac{c}{H_0}2\left[\frac{1}{(1+z_1)^{\frac{1}{2}}(1+z_2)} -
\frac{1}{(1+z_2)^{\frac{3}{2}}}\right],
\end{equation}
or
\begin{equation}
D(z_1,z_2) = \frac{c}{H_0}r(z_1,z_2)
\end{equation}
where $r(z_1,z_2)$ is the dimensionless angular
diameter distance, $c$ is the velocity of light, and $H_0$ is the
Hubble parameter.

Magnification values, $\mu$, derived using Dyer and Roeder's angular
diameter distances will be affected according to the approximation
used. For example, rays passing close to clumps or through
high-density regions will result in magnification in any
approximation. If the empty cone approximation is used, then $\mu$
will be greater than 1, and if the full cone approximation is used,
then $\mu$ will be greater than the mean magnification, $<\mu>$, since
these would be the respective values in the absence of lensing. Rays
passing through voids will have $\mu = 1$ in the empty cone
approximation (since the rays will be far from all concentrations of
matter, and will satisfy the empty cone conditions). In the full cone
approximation, $\mu < 1$ because the rays will suffer
divergence. However, the minimum value in this case will be,
(Schneider et al., 1992),
\begin{equation}
\mu_{min} = \left[\frac{D(z;\bar{\alpha}=1)}{D(z;\bar{\alpha}=0)}\right]^2.
\end{equation}
In addition, the mean value of the magnifications
derived from a large number of lines of sight through a cosmological
simulation should be close to unity in either the empty cone or the
filled beam approximation.

In the testing of our new algorithm we have used the filled beam
approximation ($\bar{\alpha}=1$) to obtain the angular diameter
distances. With variable softening, most of the rays will pass through
a slowly varying density field, justifying this choice, although the
smoothness parameter should be different from unity, and possibly
should evolve slowly with time. (Tomita, 1998b, finds, by solving the 
null-geodesic equations for a large number of pairs of light rays in 
four different cosmological $N$-body simulations, that the best value 
for $\bar{\alpha}$ is almost equal to 1, although with considerable 
dispersion.) However, the approximation that all rays should pass
far from the clumps will not be strictly true, as shear on the light
rays will be very much in evidence.

We show in Figure 4 the value of the factor $r_d r_{ds}/r_s$, where
$r_d$ is the dimensionless angular diameter distance from the observer
to the lens (here, the front face of each simulation box), $r_{ds}$ is
that between the lens and the source, and $r_s$ is that for the
observer-source, where we have taken the source to be at a redshift of
5. This factor, $r_d r_{ds}/r_s$, is used to multiply the shear
component values generated in the code, and we see that it has a peak
near $z = 0.5$ for a source redshift of 5. The curve is very steep
near $z = 0$ indicating large fractional differences between the value
of $r_d r_{ds}/r_s$ at the front and the back of simulation boxes at
late times, where considerable structure may also be present.

\begin{figure}
$$\vbox{
\psfig{figure=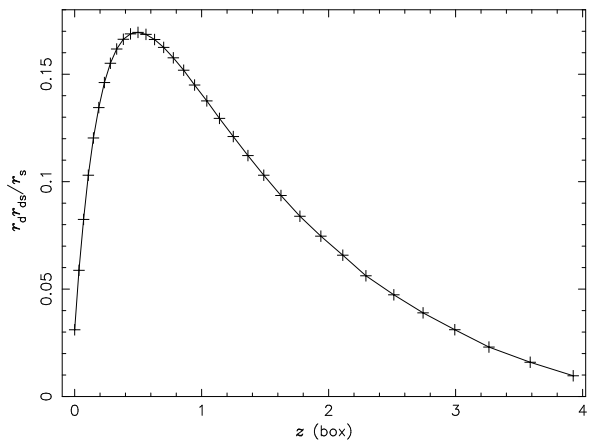,width=8.7truecm,angle=270}
}$$
\caption{The factor $r_d r_{ds}/r_s$ vs. box redshift, assuming a
source redshift of 5.}
\label{fig:angdiamA5.ps}
\end{figure}

To obtain the absolute shear component values we must also introduce
scaling factors which apply to the simulation box dimensions. The
appropriate factor is $B(1+z)^2r_d r_{ds}/r_s$, where $B=3.733 \times
10^{-9}$ for the simulation boxes we have used, which have comoving
dimensions of $100 h^{-1}$Mpc.  The $(1+z)^2$ factor occurs to convert
the comoving code units into real units. By evaluating this factor at
the front and rear faces of each simulation box, we can obtain an
estimate of the maximum error associated with projecting the mass
distribution onto a plane. In Figure 5, we plot the percentage
differences in this factor between the front and rear faces of each
simulation box, and show the results for boxes of 50$h^{-1}$Mpc,
100$h^{-1}$Mpc, and 200$h^{-1}$Mpc comoving depths. The figure clearly
indicates the possible presence of large errors when boxes are treated
as plane projections. The errors are considerable at high and low
redshifts, and, in particular, they are significant near $z=0.5$,
where the angular diameter factor $r_{d}r_{ds}/r_{s}$ is greatest. For
simulation boxes of 50$h^{-1}$Mpc the difference is 4.5\% near
$z=0.5$, for boxes of 100$h^{-1}$Mpc the difference is 9.0\%, and for
boxes of 200$h^{-1}$Mpc the difference is 16.3\%, for a source at
$z=5$.

Obviously, the front-rear differences are smallest in the smallest
boxes, (here 50$h^{-1}$Mpc comoving depth), but with small simulation
boxes there are problems in adequately representing the extent of
large-scale structure, and, as we have seen in Section 4.1, a serious
question as to whether two-dimensional shear values can be correctly
determined by considering matter out to such small radii in the
transverse direction. Even with 100$h^{-1}$Mpc boxes we have shown
that the two-dimensional shear values can be seriously in error when
only matter within the fundamental volume is included. With
200$h^{-1}$Mpc boxes, better convergence of values may be obtained
because of the larger spread of matter transverse to the line of
sight; however, the range in the angular diameter distance factors
along the line of sight is greater, introducing larger errors. To
reduce this error, it may be thought that the boxes could be divided
into a number of planes; however, this procedure would give erroneous
values for the shear because a full single period in depth is
required, as we show in Appendix B.

\begin{figure}
$$\vbox{
\psfig{figure=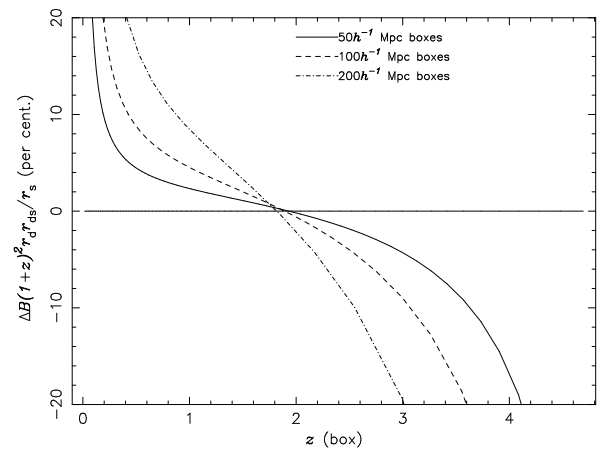,width=8.7truecm,angle=270}
}$$
\caption{The percentage difference in the multiplying factor for the
shear component values generated by the code, between the front and
back faces of each simulation box. The figure shows the differences
for simulation boxes of different comoving depths, highlighting
possible errors associated with plane projections.}
\label{fig:angdiam5comb.ps}
\end{figure}

In Section 5.3 we show that an approximation for the
magnification in weak lensing is
\begin{equation}
\mu \simeq 1 + (\psi_{11} + \psi_{22}),
\end{equation}
where the $\psi$ values are the two-dimensional
`effective lensing potentials,' derived from integrating the
three-dimensional components. Consequently, we may also find that the
errors in the shear component values, arising from ignoring the angular
diameter distance factors through the simulation boxes, enter into
calculations of the magnification.

\section{APPLICATION OF THE CODE TO LARGE-SCALE STRUCTURE SIMULATIONS}

\subsection{Brief description of the LSS simulations used}

Our three-dimensional shear code can be applied to any
three-dimensional distribution of point masses confined within a cubic
volume. Each particle may be assigned an individual mass, although
in our tests of the code, we have assumed all the particles to have
the same mass. In addition, the code allows for either a fixed
softening value for each particle, or a variable softening, dependent on
each particle's density environment.

We applied the code to the data bank of cosmological $N$-body
simulations provided by the Hydra Consortium
(http://coho.astro.uwo.ca/pub/data.html) and produced using the
`Hydra' $N$-body hydrodynamics code (Couchman, Thomas and Pearce, 1995).

Our initial tests, described here, have used individual time-slices
from these simulations using $128^3$ particles with a cold dark matter
(CDM) spectrum in an Einstein-de Sitter universe. Each time-slice has
co-moving sides of $100h^{-1}$ Mpc. Since each is generated using the
same initial conditions, we arbitrarily translate, rotate and reflect
each time-slice to prevent the formation of unrealistic correlations
of structure along the line of sight, when the boxes are linked
together. The simulations used have density parameter $\Omega_0 = 1$
and cosmological constant $\Lambda_0 = 0.$ The power spectrum shape
parameter, $\Gamma$, has been set to 0.25, as determined empirically
on cluster scales (Peacock and Dodds, 1994), and the normalisation,
$\sigma_8$, has been taken as 0.64 to reproduce the number density of
clusters (Vianna and Liddle, 1996). The dark matter particle masses
are all $1.29 \times 10^{11}h^{-1}$ solar masses.

\subsection{The choice of softening}

We have chosen a softening function for the radial distribution of
mass for each particle, such that light rays feel the existence of a
smooth mass distribution.  Our code also allows for variable
softening, so that each particle may be assigned its own
softening-scale parameter, depending on the particle number-density in
its environment. In this way, it can be used to minimise the effects
of isolated single particles, whilst the smoothed denser regions are
able to represent the form of the large-scale structure. The parameter
we have chosen to delineate the softening scale for each particle is
proportional to $l,$ where $2l$ is the radial distance to the
particle's 32nd nearest neighbour. The value of $l$ is evaluated for
every particle by applying our SPH density programme, as described in
Section 3.2, to each simulation box. 

We allow the maximum softening to be of the order of the mesh
dimension for isolated particles, which is defined by the regular grid
laid down to decompose the short- and long-range force calculations. In
this way the density values are improved, as we described in Section
2.3. This also means that individual isolated particles are unable to
strongly influence the computed shear values, in accordance with our
need to study the broad properties of the large-scale structure,
rather than the effects of individual particles. 

Our new algorithm works with the ratio of the chosen softening
(proportional to $l$) for each particle, to the maximum value
(dependent on the mesh size), so that the parameter used has a maximum
of unity.  Our method, which employs the variable softening facility,
contrasts markedly with that of other workers. As an example,
Jaroszy\'{n}ski et al. (1990), who evaluate deflections due to density
columns projected onto a plane, apply no softening function, except to
assume that all the mass within each column is effectively located at
its centre. 

In the CDM simulations we have used, the minimum values for $2l$ are
of order $10^{-3}$; e.g., for the redshift $z = 0.4986$ box, the
minimum $2l = 1.02383 \times 10^{-3}$, (equivalent to $68h^{-1}$
kpc). This is comparable to the Einstein radius, $R_E$, for a large
cluster of 1000 particles, (for which $R_E = 82h^{-1}$kpc for a lens
at $z=0.5$ and a source at $z=1$). Consequently, by setting a working
minimum value for the variable softening of $10^{-3}$ we would rarely
expect to see strong lensing due to caustics in our simulations. Also,
the radial extent of this minimum softening is of the order of
galactic dimensions, thereby providing a realistic interpretation to
the softening. 

Having justified our chosen value for a working minimum value, it
is also important to understand the sensitivity of our results to the
input softening. Figure 6 shows the distribution of magnifications due
to a single simulation box ($z = 0.4986$), and assumed source redshift
of 1. The distributions using minimum softenings of 0.001 and 0.002
are extremely close, whilst the differences are more apparent with the
minimum softenings of 0.003 and 0.004.  This is extremely helpful,
because it means that if we set the minimum softening to 0.001 in the
$z = 0$ box (equivalent to $100h^{-1}$ kpc) and to 0.002 in the $z =
1.0404$ box (also equivalent to $100h^{-1}$ kpc), then our results are
likely to be little different from results using the same minimum
softening throughout.

\begin{figure}
$$\vbox{
\psfig{figure=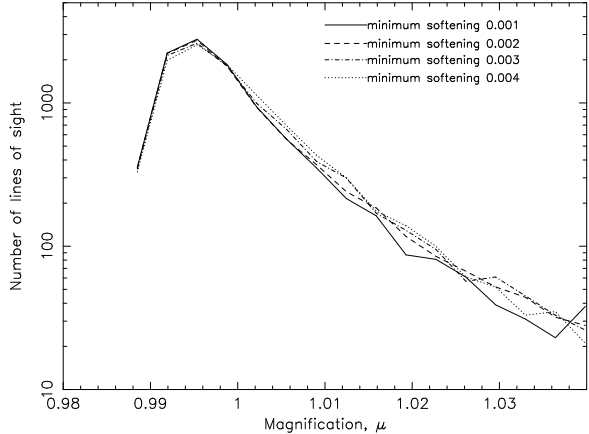,width=8.7truecm,angle=270}
}$$
\caption{Magnification distributions for various minimum softenings,
0.001, 0.002, 0.003, and 0.004 in a simulation box at $z=0.4986$, for
a source at $z=1$.}
\label{fig:comb38.ps}
\end{figure}

To highlight the sensitivity to the minimum softening, we plot in
Figure 7 the accumulating number of lines of sight having
magnifications greater than or equal to the abscissa value. As
expected, we see that the results using the smallest minimum softenings
give rise to the highest maximum magnifications.

\begin{figure}
$$\vbox{
\psfig{figure=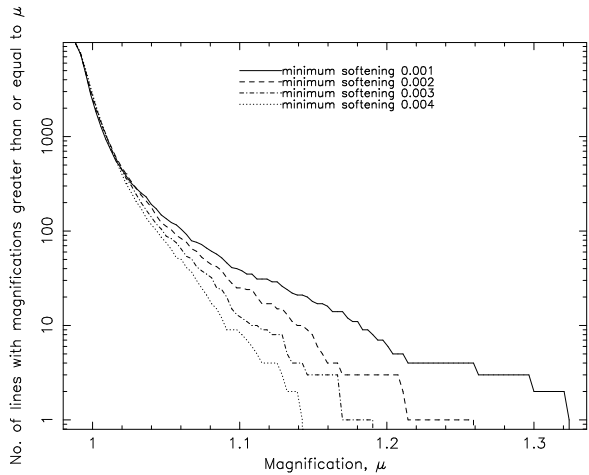,width=8.7truecm,angle=270}
}$$
\caption{The accumulating number of lines of sight for which the
magnification is greater than or equal to $\mu$. The plots show the
results of different minimum softenings in a simulation box at
$z=0.4986$, for a source at $z=1$.}
\label{fig:comb38acc.ps}
\end{figure}

\subsection{Multiple lens-plane theory for magnification
distributions}

There are two very important properties of our three-dimensional
algorithm for shear which make it eminently suitable for use within
particle simulations.

First, each simulation box is treated as a periodic system, so that
the contributions from all particles and their images are included in
the shear computations.

Second, as far as we are aware, this is the first algorithm
successfully adapted for $N$-body simulations, in which the shear
components may be evaluated at a large number of locations throughout
the extent of the box. In this way each of the selected locations may
be considered as an individual deflection site, and deflections
computed using individual angular diameter distances for each
site. Two-dimensional planar approaches, by contrast, are able to
compute only one deflection for each ray at each projection plane, and
such planes will be assumed to be at a single particular angular
diameter distance.

The first property of our algorithm gives confidence in the
shear component values computed, whilst the second property enables
us to trace the behaviour of rays throughout the full depth of each
simulation box.

To do this we construct a rectangular grid of directions
through each box. Since we are dealing with small deflections,
and are interested only in the statistics of the output values, we
consider each light ray to follow one of the lines defined by these
directions through the box. The evaluation positions are specified
along each of these lines of sight. 

The six independent second derivatives of the peculiar gravitational
potential are calculated by the code at each of the selected
evaluation positions throughout a simulation box.  We then integrate
the values in small chunks along each line, forming,
essentially, a large number of planes through each simulation box.
These integrated values form the input data to establish the elements
of the Jacobian matrix, $\cal A$, on each of the lines of sight for
each of the deflection sites.

We make use of the multiple lens-plane theory, which has been
developed by Blandford and Narayan (1986), Blandford and Kochanek 
(1987), Kovner (1987), and Schneider and Weiss (1988a, b), and
described in detail in Schneider et al. (1992). At the first deflection
site from the source we evaluate the components of the Jacobian
matrix,
\begin{equation}
\cal A = \left( \begin{array}{cc}
	1-\psi_{11}  & -\psi_{12} \\
	-\psi_{21}   & 1-\psi_{22}
	\end{array}
	\right),
\end{equation}
in which the two-dimensional `effective lensing
potentials' are obtained from the three-dimensional second derivatives
of the gravitational potential:
\begin{equation}
\psi_{ij} = \frac{D_d D_{ds}}{D_s}.\frac{2}{c^2} \int\frac{\partial^2
\phi(z)}{\partial x_i \partial x_j}dz,
\end{equation}
where $D_d$, $D_{ds}$, and $D_s$ are the angular
diameter distances from the observer to the lens, the lens to the
source, and the observer to the source, respectively. At subsequent
deflection sites we obtain the developing Jacobian matrix recursively,
since the final Jacobian for $N$ deflections is:
\begin{equation}
{\cal A} _{\rm total} = {\cal I} - \sum_{i=1}^{N}{\cal U} ^i {\cal A}_i,
\end{equation}
where $\cal I$ is the unitary matrix,
\begin{equation}
{\cal U}^i = \left( \begin{array}{cc}
	\psi_{11}^i  & \psi_{12}^i \\
	\psi_{21}^i  & \psi_{22}^i
	\end{array}
	\right)
\end{equation}
for the $i$th deflection, and each of the
intermediate Jacobian matrices can be written as
\begin{equation}
{\cal A}_j = {\cal I} - \sum_{i=1}^{j-1}\beta_{ij}{\cal U}_i{\cal A}_i,
\end{equation}
where
\begin{equation}
\beta  _{ij} = \frac{D_s}{D_{is}}\frac{D_{ij}}{D_j},
\end{equation}
in which $D_j$, $D_{is}$ and $D_{ij}$ are the angular
diameter distances to the $j$th lens, that between the $i$th lens and
the source, and that between the $i$th and $j$th lenses, respectively.

The magnification is, in general,
\begin{equation}
\mu =\left(\det \cal A \right)^{-1},
\end{equation}
so that we can assess the magnification as it
develops along a line of sight, finally computing the emergent
magnification after passage through an entire box or set of boxes.
For example, Figure 8 shows the development of the magnification
through a single isolated simulation box ($z = 0.4986$) with a chosen
source redshift of 1. The slightly different emerging magnifications arise
because of the choice of different input minimum softening values. The
figure shows an arbitrary line of sight, and we have assumed that the
redshift varies linearly through the box. It should be noted that the
magnifications derived using minimum softenings of 0.001 and 0.002 are
almost identical.

\begin{figure}
$$\vbox{
\psfig{figure=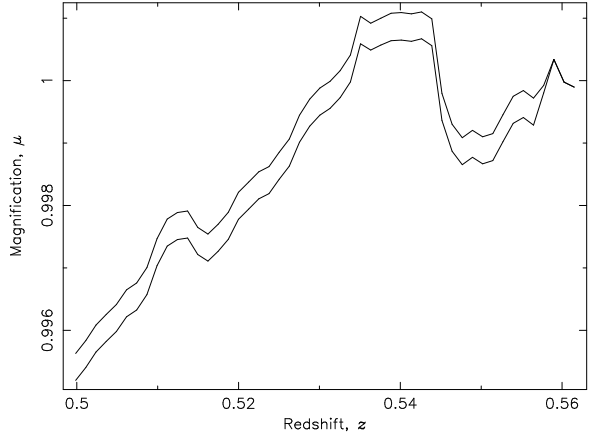,width=8.7truecm,angle=270} 
}$$
\caption{Magnification vs. redshift along an arbitrary line of sight,
using two different minimum softening values. The lower curve has a
minimum softening of 0.001, and the upper curve has a minimum
softening of 0.004.} 
\label{fig:maglinesoft}
\end{figure}

The convergence, $\kappa$, is defined by
\begin{equation}
\kappa = \frac{1}{2}(\psi_{11}+\psi_{22})
\end{equation}
from the diagonal elements of the Jacobian matrix,
and causes isotropic focussing of light rays, and so isotropic
magnification of the source. Thus, with convergence acting alone, the
image would be the same shape as, but of larger size than, the source.

The shear, $\gamma$, in each line of sight, is given by
\begin{equation}
\gamma^2 = \frac{1}{4}(\psi_{11}-\psi_{22})^2 + \psi_{12}^2.
\end{equation}
This is sometimes written in component form:
\begin{equation}
\gamma^2 = \gamma_{1}^2 + \gamma_{2}^2,
\end{equation}
where
\begin{equation}
\gamma_1 = \frac{1}{2}(\psi_{11} - \psi_{22})
\end{equation}
and
\begin{equation}
\gamma_2 = \psi_{12}.
\end{equation}
Shear introduces anisotropy, causing the image to be a different
shape, in general, from the source.

With weak lensing, and these definitions, the magnification reduces to
\begin{equation}
\mu \simeq 1 + (\psi_{11}+\psi_{22}) = 1 + 2 \kappa.
\end{equation}
In the presence of convergence and shear, a circular source becomes
elliptical in shape, with major and minor axes
\begin{equation}
a = \frac{1}{(1-\kappa -\gamma)},
\end{equation}
and
\begin{equation}
b = \frac{1}{(1-\kappa +\gamma)},
\end{equation}
so that the ellipticity, $\epsilon$, is given by
\begin{equation}
\epsilon = \frac{b}{a} = \frac{1-\kappa -\gamma}{1-\kappa +\gamma},
\end{equation}
which reduces to
\begin{equation}
\epsilon \simeq 1 - 2 \gamma
\end{equation}
in weak lensing.

Distributions and relationships amongst all these quantities can be
determined straightforwardly. As an example, Figure 9 shows the shear
and convergence in a single (assumed isolated) simulation box, ($z =
0.4986$), with a source at redshift $z_s = 1$. The minimum (variable)
softening has been set at 0.001. As expected, the shear increases
rapidly with increasing convergence, (density), until a maximum value
is reached where the minimum softening applies. Measurements of the
magnification and ellipticity show the linear dependences on $\kappa$
and $\gamma$ according to equations (21) and (25), as expected. Small
departures from linearity are apparent only at high values of $\kappa$
and $\gamma$.

\begin{figure}
$$\vbox{
\psfig{figure=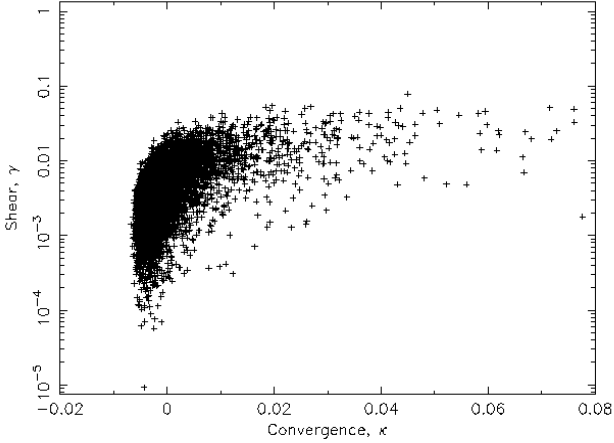,width=8.7truecm,angle=270}
}$$
\caption{Shear values vs. the convergence 
in a single simulation box, showing the rapid rise with $\kappa$, and
the flattening off when the minimum softening applies.} 
\label{fig:mag38_001gamma}
\end{figure}

Figure 10 is an example of the magnification distributions in three
different simulation boxes, which are assumed to be isolated in
space. We show the distributions for the $z = 0$ box, the $z = 0.4986$
box, and the $z = 1.0404$ box, each with minimum softenings of
0.001. A source redshift of $z_s = 2$ has been chosen. (All the
distributions have a mean magnification value of 1.)

\begin{figure}
$$\vbox{
\psfig{figure=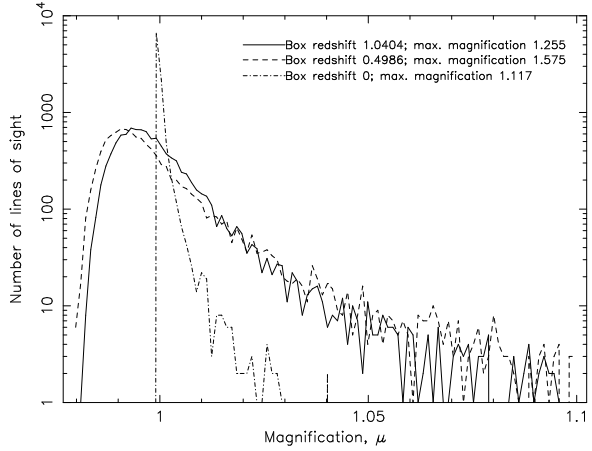,width=8.7truecm,angle=270}
}$$
\caption{A comparison of the magnification distributions for three
different simulation boxes, placed at their correct distances, with a
source at $z_s=2$.}
\label{fig:mdistcomb313849}
\end{figure}

In Figure 11 we show the results from the same three simulation boxes,
for a source at $z=2$, but here assumed to be all located at the same
position ($z=1.0404$). This allows direct comparisons between the
boxes to be made in terms of the formation of structure within
them. For example, the later boxes show higher values for the maximum
magnifications, and have shallower slopes in the distributions at the
high magnification end. The peaks in the distributions for the later
boxes occur at slightly lower magnification values, whilst all have
mean magnifications of unity, as required.

\begin{figure}
$$\vbox{
\psfig{figure=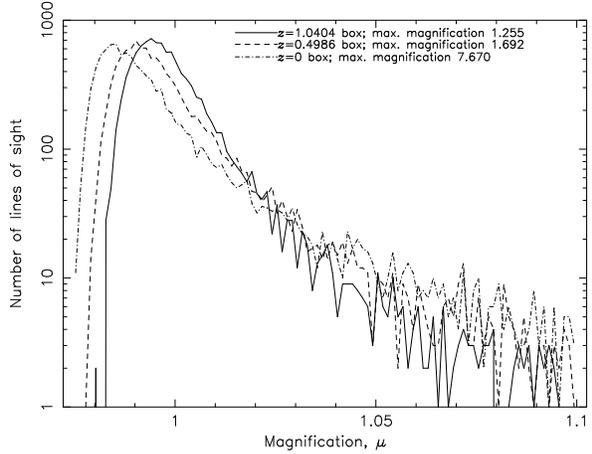,width=8.7truecm,angle=270}
}$$
\caption{A comparison of the magnification distributions in three 
different simulation boxes, all assumed to be located at the same
redshift, 1.0404.} 
\label{fig:magdistz1comb313849.ps}
\end{figure}

By a simple extension of the multiple lens-plane theory, we are now
able to take the emergent $\psi$ values from each simulation box, and
feed them into a string of subsequent boxes. In this way, we are able
to obtain all the necessary emergent parameters at $z = 0$ arising
from a source at high redshift. We shall be reporting on these results
in a future publication.  

\section{SUMMARY AND CONCLUSIONS}

In this paper we have discussed our motivations for developing a new
algorithm for use with cosmological $N$-body simulations in the study
of weak gravitational lensing. We have also described the algorithm we
have developed together with its variable softening refinement, and we
have tested the output results from three-dimensional simulations
against the Ewald (1921) summation method for the shear components. We
have described how the results from the new code can be applied to
realistic simulations by including the appropriate angular diameter
distances at every evaluation position. In this way it is very
straightforward to compute the final magnifications, source
ellipticities, shear and convergence values as a result of the passage
of light through linked simulation boxes. The main points we have
discussed are the following.

1. Appendix B rigorously shows that results from the two-dimensional
   and three-dimensional approaches to weak lensing are equivalent
   only if the mass distribution is periodic along
   the line of sight, a full period (in depth) is considered,
   and the angular diameter distances are assumed constant throughout
   the depth.

2. In order to evaluate the shear components correctly, it is
   necessary to work with the peculiar potential, which we describe in
   Appendix A. This applies equally to two-dimensional and
   three-dimensional methods.

3. The results for two-dimensional planar projections may be invalid
   if matter outside the single period plane (in directions
   orthogonal to the line of sight) is not included. Because of the
   slow convergence of the potential and shear components to
   their limiting values, it is necessary, in general, to include the
   effects of matter well beyond a single period, but depending on the
   specific mass distribution.

4. The inclusion of the appropriate angular diameter distances at
   every evaluation position within a three-dimensional realisation
   avoids errors in the shear and magnification values. The errors
   incurred by treating simulation boxes as planes may be as much as
   9\% in a single box of depth 100$h^{-1}$Mpc at a redshift of 0.5
   (where the lensing effects are greatest, for a source at
   $z=5$). (At low and high redshifts the fractional errors are
   greater.)
   
5. The output from our algorithm is the three-dimensional shear
   components evaluated at a large number of positions within a
   periodic $N$-body simulation cube. The code itself is a development
   of the standard P$^3$M algorithm which determines forces (the first
   derivatives of the potential), and the potential itself. The
   short-range part of the shear field at a point is accumulated
   directly from neighbouring particles, whilst the long-range part is
   obtained by taking a second difference of the force values. The
   computational cost of the P$^3$M method is low, being of order $N
   {\rm log}_2 N$ rather than of order $N^2$.

6. The PM calculation uses a FFT method in which the density
   distribution is smoothed, and can be well sampled by the mesh. The
   mesh potential is then obtained by FFT convolution. Errors in the
   method can be minimised by suitable adjustment of the Fourier
   components of the Green's function.

7. A key feature of the new algorithm is the facility to input a
   variable softening parameter. The feature enables particles in low
   density regions to have extended softening, so that nearby
   evaluation positions register a density rather than a complete
   absence of matter. By contrast, particles in highly clustered
   regions are assigned low softening values, and a selected minimum
   softening value is introduced to avoid singular (strong lensing)
   behaviour. The variable softening feature enables a much more
   realistic depiction of the large-scale structure within a
   simulation to be made.

8. In Appendix C we summarise the Ewald (1921) summation method,
   and develop the equations for use as a comparison with the values
   for shear obtained with our new algorithm.

9. By choosing an appropriate filter, we are able to set limits for
   the maximum errors in the computed shear values from our code. In
   the tests, the maximum errors in both the radial and the transverse
   components of the shear are about 7 per cent for the effects of a
   single particle, when compared with the values obtained using the
   Ewald formul\ae. The rms errors are less than 2 per cent, and errors
   for ensembles of particles, to which we intend to apply the code,
   are typically 0.3 per cent (rms). The errors in the trace of the
   shear matrix can be large, because the trace frequently involves
   the addition of nearly equal and opposite (but large)
   values. Individually, however, the errors in each component remain
   small.

10. We have tested the data also against the output of a completely
    different programme for the density at particle locations. There
    is good agreement for the normalised density from this
    programme when measured against the shear trace from our new
    algorithm. 

11. The output from the code can be used together with the multiple
    lens-plane theory and appropriate angular diameter distances to
    obtain values for the magnification, source ellipticity, shear and
    convergence for a large number of lines of sight as they emerge
    from a simulation box. We show a typical distribution plot for the
    emerging magnification, having given proper consideration to the
    desired minimum value for the variable softening.

12. We commend the algorithm for use in periodic $N$-body simulations
    from which the data can be manipulated to obtain emergent values
    from linked simulation cubes covering great distances. In this
    way, such a procedure also allows the comparison of results from
    different cosmologies. It is anticipated that our algorithm will
    become publicly available in enhanced form in due course.

\section*{ACKNOWLEDGMENTS}

We are indebted to the facilities supported by NSERC in Canada, and to
the Starlink minor node at the University of Sussex, for the
preparation of this paper. We thank NATO for the award of a
Collaborative Research Grant (CRG 970081) which has greatly
facilitated our interaction. AJB is sponsored by the University of
Sussex. HMPC thanks the Canadian Institute for Theoretical
Astrophysics for hospitality during the academic year 1996/97.  We
have benefited also from valuable discussions with R. L. Webster and
C. J. Fluke of the University of Melbourne, and K. Subramanian and
D. Roberts whilst they visited the University of Sussex.

\section*{REFERENCES}

\paper{Blandford R. D. \& Narayan R.}{1986}{Ap. J.}{310}{568}
\paper{Blandford R. D. \& Kochanek C. S.}{1987}{Proc. 4th Jerusalem
Winter School for Th. Physics, Dark Matter in the Universe,
ed. Bahcall J. N., Piran T. \& Weinberg S.}{Singapore, World Scientific}{p.133}
\paper{Couchman H. M. P.}{1991}{Ap. J.}{368}{L23}
\paper{Couchman H. M. P., Thomas, P. A., \& Pearce F. R.}{1995}
{Ap. J.}{452}{797}
\paper{Dyer C. C. \& Roeder R. C.}{1973}{Ap. J.}{180}{L31} 
\paper{Ewald P. P.}{1921}{Ann. Physik}{64}{253}
\paper{Fluke C. J.}{Webster R. L.}{Mortlock
D. J.}{1998a}{Preprint}
\paper{Fluke C. J.}{Webster R. L.}{Mortlock
D. J.}{1998b}{Preprint}
\paper{Gingold R. A. \& Monaghan J. J.}{1977}{MNRAS}{181}{375}
\paper{Hockney R. W. \& Eastwood J. W.}{1988}{`Computer Simulation
Using Particles'}{IOP Publishing}{ISBN 0-85274-392-0}
\paper{Holz D. E. \& Wald R. M.}{1997}{astro-ph}{9708036}{Preprint}
\paper{Jaroszy\'nski M., Park C., Paczynski B., \& Gott III J. R.}
{1990}{Ap. J.}{365}{22}
\paper{Kovner I.}{1987}{Ap. J.}{316}{52}
\paper{Marri S. \& Ferrara A.}{1998}{astro-ph}{9806053}{Preprint}
\paper{Nijboer B. R. A. \& de Wette F. W.}{1957}{Physica}{23}{309}
\paper{Peacock J. A. \& Dodds S. J.}{1994}{MNRAS}{267}{1020}
\paper{Peebles P.J.E.}{1993}{`Principles of Physical Cosmology'}{Princeton
University Press}{ISBN 0-691-07428-3}
\paper{Premadi P., Martel H. \& Matzner R.}{1998}{Ap. J.}{493}{10}
\paper{Refsdal S.}{1970}{Ap. J.}{159}{357}
\paper{Schneider P., Ehlers J., \& Falco E. E.}{1992}
{`Gravitational Lenses'}{Springer-Verlag}{ISBN 0-387-97070-3}
\paper{Schneider P. \& Weiss A.}{1998a}{Ap. J.}{327}{526}
\paper{Schneider P. \& Weiss A.}{1998b}{Ap. J.}{330}{1}
\paper{Tomita K.}{1998a}{astro-ph}{9806003}{Preprint}
\paper{Tomita K.}{1998b}{astro-ph}{9806047}{Preprint}
\paper{Vianna P. T. P., \& Liddle A. R.}{1996}{MNRAS}{281}{323}
\paper{Wambsganss J.}{1990}{Thesis of Max-Planck-Institut F\"{u}r
Physik und Astrophysik}{Ref. MPA}{550}
\paper{Wambsganss J., Cen R., Xu G. \& Ostriker
J. P.}{1997}{Ap. J.}{475}{L81}
\paper{Wambsganss J., Cen R., \& Ostriker J.}{1998}{Ap. J.}{494}{29}

\onecolumn

\appendix
\section{PERIODICITY AND THE PECULIAR POTENTIAL}
 
A common method for modelling a section of the universe is to consider
a distribution of masses within a triply periodic cube. In this
appendix we examine how the peculiar potential in this model relates
to that in a universe with large-scale homogeneity.
 
\subsection{The peculiar potential}
 
In Friedmann--Robertson--Walker models when considering the growth of
perturbations, gravitational lensing by cosmic structure, etc., we are
interested in deviations from homogeneity. The quantity of interest in
this case is the {\em peculiar} potential. This is derived in the
usual way by writing the equation of motion,
\begin{equation}
{d^2 \vec{r}\over d t^2} = -\nabla_{\vec{r}}\Phi,
\end{equation}
for matter at position $\vec{r}$, in terms of a comoving coordinate
$\vec{r}=a \vecx$, where $a$ is the expansion scale factor. The
peculiar potential---the source for deviations from homogeneity---is 
\begin{equation}
\phi=\Phi + \fracnum{1}{2} a\ddot a x^2.
\end{equation}
The rate of change of expansion velocity, $\ddot a$, is determined by
the mean density of matter, $\bar\rho$, leading to the familiar
results 
\begin{eqnarray}
\nabla^2\phi &=& 4 \pi G a^2 (\rho - \bar\rho) \label{eq:nablapec} \\
\noalign{\noindent and}
\phi&=&\Phi - \fracnum{2}{3}\pi G a^2 \bar\rho x^2
\end{eqnarray}
\eqnote{eq:nablapec}
(for full details see, for example, Peebles,~1993). The peculiar
potential arising from Poisson's equation~(\ref{eq:nablapec})
corresponds to a system with zero net mass on large scales. 
 
Consider now a model universe of masses periodic in a cube of side
$L$. The forces generated by the matter distribution satisfy
$\vec{F}(\vec{x} + L\vec{n})=\vec{F}(\vec{x})$ where \vec{n} is an
integer triple. Thus integrals, $\int_{L^2} \vec{F}.d\vec{S}$, where
$\vec{S}$ is an outward normal, taken over opposite faces of the cube,
sum to zero, and $\int_S \vec{F}.d \vec{S}$ over the surface of the
cube vanishes. The divergence theorem and Poisson's equation,
$\nabla^2\phi=4\pi 
G\rho^\prime$, then imply  
\begin{equation}
 4\pi G \int_{L^3} \rho^\prime\, dV =  \int_{L^3}\nabla^2\phi\,
d V = - \int_{L^3}\nabla.\vec{F}\,d V = -\int_S \vec{F}.d \vec{S} =
0,\label{eq:mzper} 
\end{equation}
\eqnote{eq:mzper}
and hence the total mass in the system is zero.
 
This result is a consequence of solving Poisson's equation in a
periodic cube. We see that the real result for the shear, obtained
from the second derivatives of the peculiar potential, $\phi$, is
related to the na\"{\i}ve result based on the full gravitational
potential, $\Phi$, through the use of $\rho^\prime=\rho - \bar\rho$.  
 
Note that the zero mean density implicit in
equation~(\ref{eq:nablapec}) is a result of the coordinate
transformation; that this transformation is well motivated is a
reflection of the large-scale homogeneity of the universe. The zero
mean density in equation~(\ref{eq:mzper}), on the other hand, is
simply a result of the imposed periodicity. The two views converge
only for a periodic cube sufficiently large that the amplitude of the
first few discrete Fourier modes are small enough that they describe a
smooth transition to a zero mean value and homogeneity.
\subsection{The peculiar potential in a periodic
system\label{subsec:pecpot}}
\eqnote{subsec:pecpot}
 
We begin by demonstrating a useful result relating the
Fourier transforms of continuous functions and the Fourier
coefficients of the periodic functions constructed from them. First,
as a convenient mechanism for translating between Fourier
representations of continuous and periodic functions,
we define the three-dimensional comb
$\comb(x)=\sum_\vecn\delta(\vecx-\vecn)$, where $\vecn$ is an integer
triple (see, for example, Hockney \& Eastwood~1988). The Fourier
transform of this function is the comb $\comb(\veck/2\pi)$. 
 
Consider the 
function, with period $L$, constructed by periodically repeating $f$: 
\begin{eqnarray}
f^\dagger(\vecx) &=& \comb(\vecx/L) * f(\vecx) \label{eq:fconv}\\
                 &=& \sum_\vecn f(\vecx-\vecn L).
\end{eqnarray}
\eqnote{eq:fconv}
Applying the Fourier convolution theorem to equation~(\ref{eq:fconv}) we
see immediately that the Fourier transform of the periodic function,
$f^\dagger$, is  
\begin{equation}
\widetilde f^\dagger_\veck = \comb\left({\veck\over L/2\pi}\right) \widetilde
f(\veck), \label{eq:ctsper} 
\end{equation}
\eqnote{eq:ctsper}
where $\widetilde f$ is the continuous Fourier transform of $f$.
 
Equation~(\ref{eq:ctsper}) gives the Fourier series representation
of the periodic function directly and demonstrates the familiar
result that the Fourier coefficients of the continuous periodic
function, obtained by accumulating repeats of the continuous
function (here $\sum_\vec{n} f(\vecx-\vec{n} L)$ obtained from $f$),
are the same as the Fourier components of the continuous function at
wavenumbers $\veck=2\pi\vecl/L$. This result is the analogue in real
space of aliasing in Fourier space.
If the continuous function is zero outside the
fundamental cell the periodic representation is simply obtained by
tiling space with repeats of the fundamental cell. (This is the
real-space analogue of a band-limited function.) 
 
Consider now $N$ particles distributed in a cube of side $L$. Let the
position of particle $i$ be $\vecxi$. The density is then
\begin{equation}
\rho(\vecx)=\sum_j m_j\delta(\vecx - \vecxj),\label{eq:nopermass}
\end{equation}
\eqnote{eq:nopermass}
without, for the moment, requiring zero total mass. Where necessary,
we can consider a periodic density distribution constructed by tiling
space with periodic repeats of the distribution 
in the fundamental cube:
\begin{equation}
\rho^\prime(\vecx)=\comb(\vecx/L) * \rho(\vecx).\label{eq:permass}
\end{equation}
\eqnote{eq:permass}
 
The gravitational potential at a point $\vecx$ in the periodic system
is 
\begin{equation}
\phi(\vecx)=G \comb(\vecx/L) * \rho(\vecx) * \varphi(\vecx),
\label{eq:intconv} 
\end{equation}
\eqnote{eq:intconv}
where $\varphi$ is the pairwise `interaction' potential (or Green's
function). Equation~(\ref{eq:intconv}) may be interpreted in two
ways. We may consider a periodic distribution of matter as in
equation~(\ref{eq:permass}) convolved with the regular interaction
potential. Alternatively, we can restrict attention to the matter in
the fundamental zone and consider a modified interaction potential
\begin{eqnarray}
\varphi^\dagger(\vecx) &=& \comb(\vecx/L)  * \varphi(\vecx) \\
                       &=& \sum_\vecn \varphi(\vecx - \vecn
                       L).\label{eq:phidag1} 
\end{eqnarray}
\eqnote{eq:phidag1}
Both of these interpretations will be useful. 

Using the result in equation~(\ref{eq:intconv}), the Fourier
series coefficients of $\phi$ are, for $\veck=2\pi\vec{l}/L$ and
$\vecl$ an integer triple,
\begin{equation}
\widetilde\phi_\veck=G\, \widetilde\rho(\veck)
\widetilde\varphi(\veck),\label{eq:phifs}
\end{equation}
\eqnote{eq:phifs}
where $\widetilde\rho(\veck)=\sum_j m_j {\rm e}^{-i\veck.\vecxj}$ is
the continuous Fourier transform of the particle distribution in the
fundamental zone (equation~(\ref{eq:nopermass})) and
$\widetilde\varphi(\veck)$ is the continuous Fourier transform of the
interaction potential.
 
Requiring the mean mass to be zero is equivalent to subtracting a
component equal to $\sum_i m_i/L^3$ from the density in
equation~(\ref{eq:nopermass}), or setting $\widetilde\phi_{\veck=0} =
0$ in equation~(\ref{eq:phifs}). With this modification
equation~(\ref{eq:intconv}) becomes:
\begin{eqnarray}
\phi(\vecx) &=& G \comb(\vecx) * \sum_j m_j \left[ \varphi(\vecx - \vecxj) -
{1\over L^3} \int_{L^3} \varphi(\vecx -\vecx^\prime)\, d^3
\vecx^\prime\right]\\
            &=& G\left[\sum_\jn m_j \varphi(\vecx-\vecxjn) -{\sum_j m_j\over
L^3} \sum_\vecn \int_{L^3} \varphi(\vecx-\vecn L -\vecx^\prime)\, d^3 
\vecx^\prime\right]\\
            &=& G \sum_j m_j\left[\sum_\vecn \varphi(\vecx-\vecxjn) -
{1\over L^3} \int\varphi(\vecx^\prime)\, d^3
\vecx^\prime\right]\label{eq:perzerop}\\ 
            &=& G \sum_\jn m_j \left[ \varphi(\vecx-\vecxjn) - 
{1\over L^3} \int_{L^3} \varphi(\vecx^\prime - \vecn L)\, d^3
\vecx^\prime\right],\label{eq:perzero}
\end{eqnarray}
\eqnote{eq:perzerop}
\eqnote{eq:perzero}
where $\vecxjn=\vecxj + \vecn L$ and $\jn$ labels the image of the $j$-th
particle in the image cell $\vecn$. We may write
equation~(\ref{eq:perzero}) as
\begin{equation}
\phi(\vecx) = G \rho(\vecx) * \varphi^\dagger(\vecx),
\end{equation}
where we now have for the density
\begin{equation} 
\rho(\vecx)=\sum_j m_j \left[ \delta(\vecx - \vecxj) - {1\over
L^3}\right],\label{eq:fulldens}
\end{equation}
\eqnote{eq:fulldens}
and equation~(\ref{eq:phidag1}) is now 
\begin{equation} 
\varphi^\dagger(\vecx) = \sum_\vecn\left[ \varphi(\vecx-\vecn L) -
{1\over L^3} \int_{L^3} \varphi(\vecx^\prime - \vecn L)\, d^3
\vecx^\prime\right].\label{eq:fullvdag}
\end{equation}
\eqnote{eq:fullvdag}
It is unnecessary for the mean value of {\em both} $\rho$ and
$\varphi$ to be zero. The form for the modified potential is
convenient, however, since, for a potential which is Coulombic at large
scales, with $\varphi\sim 1/r$, the form in equation
~(\ref{eq:fullvdag}) is convergent whereas the form in
equation~(\ref{eq:phidag1}) is not: nor does the integral in
equation~(\ref{eq:perzerop}) converge. The Fourier synthesis 
of the peculiar potential is
\begin{equation}
\phi(\vecx)={G\over L^3} \sum_{\veck\ne 0} e^{i\veck.\vecx}
\widetilde\rho(\veck)\, \widetilde\varphi(\veck).\label{eq:fpec}
\end{equation}
\eqnote{eq:fpec}

\subsection{Particle softening\label{subsec:soft}}
\eqnote{subsec:soft}
 
The delta function in equation~(\ref{eq:nopermass}) may be replaced by
any compact even function, with volume integral equal to unity, to
represent a distribution of softened particles. (The softening will be
considered fixed for the present to permit Fourier analysis.) However,
since it is only the {\it interaction} of the particles via the
gravitational field which is relevant, we may, equivalently, describe
any particle softening by modifying the pairwise (Coulombic)
potential, $\varphi$, at small separations. 
 
From equation~(\ref{eq:fpec}) we can immediately write
\begin{equation}
\nabla^2\phi = - {G\over L^3}\sum_{\veck\ne 0}
e^{i\veck.\vecx}\widetilde\rho(\veck)\, k^2
\widetilde\varphi(\veck).\label{eq:poiss} 
\end{equation}
\eqnote{eq:poiss}
 
For a Coulombic interaction potential, $\varphi(r)=-1/r$, we have
$\widetilde\varphi(k)=-4\pi/k^2$. If we set
$\widetilde\varphi(k)=-4\pi\widetilde S(k)/k^2$ with $\widetilde S(0)
= 1$, then $S(R)$ describes the departure of the interaction potential
from Coulombic at small scales and, as we will see, plays the role of
a particle softening. Equation~(\ref{eq:poiss}) becomes
\begin{eqnarray}
\nabla^2\phi&=&{4\pi G\over L^3} \sum_{\veck\ne0} e^{i\veck.\vecx}
\widetilde\rho(\veck) \widetilde S(\veck)\\
            &=&{4\pi G} \rho(\vecx) * S(\vecx)\\
            &=&{4\pi G} \rho_S
\end{eqnarray} 
which is Poisson's equation for a distribution of softened point
charges;
\begin{equation}
\rho_S(\vecx) ={4\pi G} \sum_j \left[ S(\vecx-\vecxj)-{1\over L^3}\right].
\end{equation}
 
Note, that if we wanted the force on a softened particle from the
distribution of softened particles (rather than merely sampling the
density field at a point), the appropriate interaction potential in
Fourier space would be $\widetilde\varphi=-4\pi \widetilde
S^2(k)/k^2$.

\section{EQUIVALENCE OF 2-D AND 3-D SHEAR CALCULATIONS}
 
\subsection{General}
 
It is frequently assumed that for the purposes of determining deflections
and shearing of light a three-dimensional mass distribution
may be represented by a plane projection of the density.
In particular, many workers in the field of gravitational lensing
treat cosmological $N$-body simulation cubes as collapsed planes. In
this Appendix we investigate this assumption and show under what
conditions the result holds.
 
The result is approximate because of the need to apply the appropriate
angular diameter distances at every deflection site (or evaluation
position). In our derivation we assume that these factors are constant
along the line of sight through the projected volume, whereas in
practice they will vary slightly through the simulation volume.  (The
technique developed in this work applies the angular diameter 
distances at every evaluation position within each three-dimensional
realisation, and evaluates the shear components at many locations
within each to enable a complete description of the shearing of a
light ray during its travel through the simulation.)
 
We will show that computations based on two-dimensional (planar)
projections of three-dimensional (periodic) simulations are adequate
provided: (a) the mass distribution is periodic along the line of
sight, and a single (full) period is included in the projection; (b)
proper account is taken of the full transverse extent of matter,
(which should normally be assumed to be periodic, unless strong
lensing by matter limited in extent is being considered); by ignoring
this requirement, it is likely that the convergence of deflection
angles and shear components to their limiting values will not be
achieved; (c) the net zero mass requirement is
adopted.
 
Consider the peculiar potential, $\phi(\vecx)$, in a three-dimensional
periodic system as given above in equation~(\ref{eq:perzero}):
\begin{equation}
\phi(\vecx)= G\sum_\jn m_j \left[ \varphi(\vecx -\vecxjn) - 
{1\over L^3}\int_{L^3} \varphi(\vecx^\prime -\vecn L)\,
d^3\vecx^\prime \right].\label{eq:ppotperb}
\end{equation}
\eqnote{eq:ppotperb}

As discussed in Section~\ref{subsec:pecpot}, the integral in
equation~(\ref{eq:ppotperb}) is finite and the sum over 
$\vecn$ converges. Although we will be considering derivatives of the
potential---in which case the second, constant, term drops out---it is useful
to have a rigorous convergent expression for the peculiar potential in
a periodic system.
 
We are interested in obtaining the integrated shear along a
line-of-sight over one period: $\int_L
\partial^2\phi(\vecx)/\partial x_i\partial x_j\, dz$. We will begin by
integrating the peculiar potential over one period:
$\int_L\phi(\vecx)\, dz$. 
 
In the following
we will split vectors over three 
dimensions into a two-dimensional component perpendicular to the
line-of-sight and a component along the line-of-sight, here taken to
be in the $z$ direction. A superscript asterisk is used to denote
2-dimensional quantities, e.g., $\vecx=(\vecX,z)$. Then the
two-dimensional potential is
\def\hm{\hskip -8pt}
\begin{eqnarray}
\phi^*(\vecX)\hm&=&\hm\int_L \phi(\vecx)\, dz  \nonumber \\
  \hm&=&\hm G  \sum_{\jn^*} m_j \sum_s \left[\int_L dz\,
  \varphi(\vecX-\vecXjn, z-z_{js}) 
-{1\over L^2}\int_L \int_{L^2} \varphi({\vecX}^\prime-\vecn^* L,
z^\prime - s L)\,d^2 {\vecX}^\prime dz^\prime\right] \nonumber \\
  \hm&=&\hm G \sum_{\jn^*} m_j\left[ \int dz\, \varphi(\vecX-\vecXjn,
z-z_j) 
 -{1\over L^2} \int\!\!\int_{L^2} \varphi({\vecX}^\prime - \vecn^* L, z)\,
 d^2 {\vecX}^\prime dz^\prime\right] \nonumber \\
 \hm&=&\hm G\sum_{\jn^*} m_j\left[\varphi^*(\vecX-\vecXjn)  
-{1\over L^2} \int_{L^2} \varphi^*({\vecX}^\prime - 
{\vecn}^* L)\,d^2 {\vecX}^\prime\right],\label{eq:ppot2d}
\end{eqnarray}
\eqnote{eq:ppot2d}
where 
\begin{equation}
\varphi^*({\vecX})=\int \varphi({\vecX}, z)\, dz. 
\end{equation}
Equation~(\ref{eq:ppot2d}) is the 2-dimensional analogue of
equation~(\ref{eq:ppotperb}), with the interaction
potential, $\varphi$ 
replaced by $\varphi^*$. Thus the three-dimensional peculiar potential 
integrated over one period 
$L$ in one dimension, gives the same result as that obtained
from the projected (surface) density of particles with a 2-dimensional
interaction potential arising from the projection from \underbar{$-\infty$ to
$\infty$} of the 3-dimensional 
interaction potential.  The
corresponding results for the shear components,
$\partial^2\phi^*/\partial x_i\partial x_j$, $x_i, x_j\ne z$,
follow directly.
 
\subsection{Special Case of a Coulombic Potential}
 
For the case of a Coulombic potential in 3-dimensions, $\varphi=-1/r$,
where $r^2={\vecX}^2+z^2$, $\varphi^*({\vecX})=-\int dz / r$
diverges. Consider the two-dimensional potential over a finite range
in z, $0 \leq z \leq M$,
\begin{eqnarray}
\varphi_M^*({\vecX})&=&2\int_0^M\!\! \varphi dz \nonumber \\
 &=& -2\int_0^M {dz\over\sqrt{{\vecX}^2+z^2}} \nonumber \\
&=& 2\ln |\,\vecX| - 2 \ln\Big(M + \sqrt{ {\vecX}^2 + M^2}\, \Big).
\end{eqnarray}
From equation~(\ref{eq:ppot2d}) we can then write
\begin{eqnarray}
\phi^*(\vecX)&=&\lim_{M\rightarrow\infty} \left\{ G\sum_\jn m_j 
\left[\varphi_M^*(\vecX-\vecXjn) 
 -  { 1 \over L^2}  
\int_{L^2} \varphi_M^*({\vecX}^\prime - {\vecn}^{*\prime} L)\,d^2
{\vecX}^\prime\right]\right\} \nonumber \\
  &=&G\sum_\jn m_j \left[2\ln|\vecX-\vecXjn| - {1\over L^2}  
\int_{L^2} 2\ln|{\vecX}^\prime - {\vecn}^{*\prime} L|\,d^2
{\vecX}^\prime\right].
\end{eqnarray}
Thus, for $\varphi(r)=-1/r$, the appropriate two-dimensional potential is
$\varphi^*(r^*)= 2\ln(r^*)$ as expected.

\subsection{Softening}
 
The discussion above applies to any suitably well-behaved interaction
potential, $\varphi$. In particular, consider the case of a
distribution of softened particles. This may be described in terms of
an interaction potential which is Coulombic at large scales but which
falls below $1/r$ at small scales. It is most convenient to derive the
appropriate $\varphi^*$ in Fourier space. We may set
$\widetilde\varphi(k)= -4\pi\widetilde S^2(k)/k^2$ with $\widetilde
S(0)=1$ (see section~(\ref{subsec:soft}). The required two-dimensional
Fourier transform is then 
\begin{eqnarray}
\widetilde\varphi^*(k^*)&=&\int e^{i\veck^*.\vecx^*} d^2 \vecx^* \int
\phi(\vecx)\, d z \nonumber \\
                    &=&\widetilde\phi({\veck}^*,0).
\end{eqnarray}
 
Thus, for a given softening function $S$, the appropriate function
$\varphi^*$ may be found, although an analytic solution may not be
possible especially in view of the notorious difficulty of
two-dimensional Fourier integrals.
 
\section{THE EWALD SUMMATION METHOD}
 
In this appendix we turn to the numerical evaluation of sums for the
potential and its derivatives in a periodic system such as those in
equation~(\ref{eq:ppotperb}). For a Coulombic potential $\int
\varphi(\vecx)\, d^3 \vecx$ is divergent and the potential is only well
defined (and convergent) if there is a uniform negative mass density
to cancel the distribution of positive mass particles. Even if this
condition is met the sum for the potential is only slowly convergent and
difficult to compute numerically.  Ewald (1921) proposed a method for
computing such sums in the context of calculating lattice potentials
of ionic crystals. The electrostatic problem suffers from exactly the
same numerical difficulties as the gravitational problem (the pairwise
potential in each case is Coulombic), and it is well known that 
na\"{\i}vely summing over images of the fundamental cell gives an
order-dependent result. Note that the requirement for zero total
mass is the same as the requirement in calculating crystal energies
that the total charge be zero.
We derive below Ewald's method as it is
applied to the problem of computing the gravitational potential and
give expressions for the first and second derivatives of the
potential---respectively the force and shear---and the total
potential. We also demonstrate the relationship of the P$^3$M
technique to the Ewald method.
 
\subsection{The Ewald method}
 
Consider again a system of $N$ particles in a cube of side $L$. The
density is given by equation~(\ref{eq:fulldens}):
\begin{equation}
\rho(\vecx)=\sum_j m_j\left[\delta(\vecx - \vecxj) - {1\over L^3}
\right].\label{eq:ewdens} 
\end{equation}
\eqnote{eq:ewdens}
The second term on the right hand side of equation~(\ref{eq:ewdens})
makes the mean 
density zero as required for the existence of a
solution to Poisson's equation. As noted in section~\ref{subsec:soft},
the delta function in equation~(\ref{eq:ewdens}) may
be replaced by any compact even function, with volume integral equal
to unity, to represent a distribution of particles with fixed
softening. Alternatively, and equivalently, the interaction potential
may be suitably modified.

The gravitational potential at a point $\vecx$ within the cube is
given by equation~(\ref{eq:perzero})
\begin{equation}
\phi(\vecx)= G \sum_\jn m_j \left[ \varphi(\vecx-\vecxjn) - 
{1\over L^3} \int_{L^3} \varphi(\vecx^\prime - \vecn L)\, d^3
\vecx^\prime\right],\label{eq:ewpec}
\end{equation}
\eqnote{eq:ewpec}
where the notation is as in section~\ref{subsec:pecpot}.
Note that evaluating equation~(\ref{eq:ewpec}) at a particle 
position, $\vecxi$, will include the self energy of particle $i$. 

The sum in equation~(\ref{eq:ewpec}) converges very slowly 
and is ill conditioned for numerical computation.
Ewald~(1921) proposed
splitting the Coulombic potential into two components,
\begin{equation}
\varphi(R)=\varphi_1(R) + \varphi_2(R),\label{eq:ewsplit}
\end{equation}
\eqnote{eq:ewsplit}
where the functional form of the split is chosen so that the first
component is dominated by quickly converging local contributions and
the second contains the relatively smooth long range components of the
field. The attenuation of high frequencies in the second component
ensures rapid convergence of the corresponding sum when recast as a
Fourier series. Ewald 
proposed taking $\varphi_1=-{\rm erfc}(\eta R)/R$, 
$\varphi_2(R)=-1/R-\varphi_1(R)=-{\rm erf}(\eta R)/R$, where erf and
erfc are the error and complementary error functions respectively. The
parameter $\eta$ is chosen to optimize convergence of the resulting
real- and Fourier-space sums. For the moment we will not specify the
functional form and will continue with the description in
equation~(\ref{eq:ewsplit}). 
 
If we ignore for the moment the mean contribution in
equation~(\ref{eq:ewpec}), we can write the potential as 
\begin{equation}
\phi(\vecx)=G\sum_\jn m_j\varphi_1(\vecx-\vecxjn)
           +{G\over L^3}\sum_\veck e^{i\veck.\vecx}
\widetilde\Phi_{2\,\veck}.\label{eq:ewpecsplit}
\end{equation}
\eqnote{eq:ewpecsplit}
The second term is a Fourier series sum over $\veck=2\pi \vecl/L$, 
$\vecl$ an integer triple, because of the periodicity of the
system. Referring to the result in equation~(\ref{eq:phifs}) we see that
the Fourier components  $\widetilde\Phi_{2\,\veck}$ are given by 
\begin{equation}
\widetilde\Phi_{2\,\veck} =
\widetilde\rho^\prime(\veck)\widetilde\varphi_2(\veck),\label{eq:ewpecfor} 
\end{equation}
\eqnote{eq:ewpecfor}
where $\rho^\prime(\vecx)=\sum_j\delta(\vecx-\vecxj)$.
This is completely equivalent to the results of
subsection~\ref{subsec:pecpot} but with $\widetilde\varphi$ replaced by
$\widetilde\varphi_2$, the only difference being that there will be a
much larger effective softening for $\varphi_2$.
 
\smallskip
 
We will now ensure that the mean density is 
zero. It is not sufficient simply to set
$\widetilde\Phi_{2,\veck=0}=0$ in 
equation~(\ref{eq:ewpecsplit}) as part of the mean value of the field is
contained in the 
first (real-space) term. The required potential is
\begin{equation}
\phi(\vecx)=G\sum_{\jn}m_j\varphi_1(\vecx-\vecxjn)
           -{G\sum_j m_j\over L^3}\,\widetilde\varphi_1(\veck=0)
           +{G\over L^3}\sum_{\veck\ne0} e^{i\veck.\vecx}
\widetilde\Phi_{2\,\veck}.\label{eq:ewpecpot}
\end{equation}
\eqnote{eq:ewpecpot}
 
The continuous Fourier transform of the potential in
equation~(\ref{eq:ewpecpot}) gives, after straightforward
manipulation: 
\begin{equation}
\widetilde\phi(\veck)=G\widetilde\varphi(\veck) \left[
  \widetilde\rho^{\,\prime}(\veck)-\textstyle{\sum_j} m_j \delta(\veck)\right]
  =G\widetilde\varphi(\veck)\widetilde\rho(\veck),
\end{equation}
and thus, as required, the potential is the convolution of the total density,
$\rho$, with the interaction potential, $\varphi$. Of course, for 
$\varphi(R)=-1/R\Rightarrow\widetilde\varphi(k)=-4\pi/k^2$ and we
recover Poisson's equation in Fourier space.
 
Substituting for $\widetilde\rho^{\,\prime}(k)$ in
equation~(\ref{eq:ewpecfor}) and then 
using the result for $\widetilde\Phi_{2\,\veck}$ in
equation~(\ref{eq:ewpecpot}) gives the final result
\begin{equation}
\phi(\vecx)=G\sum_{\jn}m_j\varphi_1(\vecx-\vecxjn)
           -{G\sum_j m_j\over L^3}\,\widetilde\varphi_1(\veck=0) 
           +{G\over L^3}\sum_{j,\veck\ne0} e^{i\veck.(\vecx-\vecxj)} m_j\,
\widetilde\varphi_2(k).\label{eq:ewfin}
\end{equation}
\eqnote{eq:ewfin}
The potentials $\varphi_1$ and $\varphi_2$ can be chosen so that both
sums in equation~(\ref{eq:ewfin}) converge rapidly.

\medskip
 
For a given split of $\varphi$ in equation~(\ref{eq:ewsplit}) we can calculate
$\widetilde\varphi_2(k)$ and hence efficiently calculate the potential
at a point for a given particle distribution. From the expression in
equation~(\ref{eq:ewfin}) we may also derive expressions with similar
desirable 
convergence properties for derivatives of the potential, such as the
force and tidal field at $\vecx$:
\begin{eqnarray}
& &{\partial^s\phi\over\partial x_{\mu_1} \partial x_{\mu_2} ... \partial
x_{\mu_s}}=\qquad\qquad\nonumber\\ 
& &\hskip0.1inG\Bigg\{ \sum_{\jn} m_j
{\partial^s\varphi_1(r)\over \partial r_{\mu_1} \partial
r_{\mu_2}...\partial r_{\mu_s}}\Bigg\vert_{\vec{r}=\vecx-\vecxjn} \!\!\!\!  +\, 
{1\over L^3}\!
\sum_{j,\veck\ne0}i^s k_{\mu_1} k_{\mu_2} ... k_{\mu_s}
e^{i\veck.(\vecx-\vecxj)} m_j \widetilde\varphi_2(k)\Bigg\}.\label{eq:ewpart}
\end{eqnarray}
\eqnote{eq:ewpart}
For the first and second derivatives of $\varphi_1(r)$ we have
$\partial\varphi_1/\partial r_\mu $
$= (\varphi_1^\prime/r) r_\mu$ 
and $\partial^2\varphi_1/\partial r_\mu\partial r_\nu $
$=[(\varphi_1^\prime/r)^\prime/r] r_\mu r_\nu +$ $(\varphi_1^\prime/r)
\delta_{\mu\nu}$.

\noindent From equation~(\ref{eq:ewpart}) we can immediately write
\begin{equation}
\nabla^2\phi=G\Bigg\{\sum_\jn m_j\nabla^2\varphi_1(\vecx-\vecxjn)
- {1\over L^3}\sum_{j,\veck\ne0}
k^2e^{i\veck.(\vecx-\vecxj)}m_j\widetilde\varphi_2(k)\Bigg\}.
\end{equation}
We can recast the first term as a Fourier series, since it is
periodic; as before the coefficients for the Fourier sum are the same
as those for the continuous transform. Using this result we obtain
\begin{equation}
\nabla^2\phi={G\over L^3}\Bigg\{\sum_{j,\veck} m_j
e^{i\veck.(\vecx-\vecxj)} \int
e^{-i\veck.\vecx^\prime}\nabla^2\varphi_1\,d^3\vecx^\prime - \sum_{j,\veck\ne0}
k^2e^{i\veck.(\vecx-\vecxj)}m_j\widetilde\varphi_2(k)\Bigg\},\label{eq:ewpoiss1} 
\end{equation}
\eqnote{eq:ewpoiss1}
where the integral is over all space. Note that the first sum includes
the mean $\veck=0$ component. Equation~(\ref{eq:ewpoiss1}) leads directly to
\begin{equation}
\nabla^2\phi=-{G\over L^3}\Bigg\{\sum_{j,\veck}
e^{i\veck.(\vecx-\vecxj)} m_j k^2\widetilde\varphi -
\lim_{k\rightarrow0}\big[k^2\widetilde\varphi_2(k)\big]\sum_j
m_j\Bigg\}.\label{eq:ewpoiss2}
\end{equation}
\eqnote{eq:ewpoiss2}
 
If the interaction potential is Coulombic on large scales,
$\varphi\sim-1/R$, then $\lim_{k\rightarrow0}k^2\widetilde\varphi_2 =
-4\pi$, provided that $\varphi_1/\varphi_2\rightarrow0$ sufficiently
rapidly with increasing $R$ (this condition is satisfied for any
practical splitting choice). Equation~(\ref{eq:ewpoiss2}) then gives
\begin{equation}
\nabla^2\phi=4\pi G \rho,
\end{equation}
{\em including the negative mean density} as in equation~(\ref{eq:ewdens}). 
 
\smallskip

The total potential energy of the $N$ particles in the box including
interactions with all images is
\begin{equation}
U={G\over 2}\sum_{\scriptstyle i,\jn\atop\scriptstyle \jn\ne i} 
     m_i m_j \varphi(\vecxi-\vecxjn)\label{eq:ewtotpot}
\end{equation}
\eqnote{eq:ewtotpot}
(ignoring again temporarily the mean contribution to the
peculiar potential in equation~(\ref{eq:ewpec})).
We can make use of the same splitting in equation~(\ref{eq:ewsplit})
and the results 
that follow from it for the potential, provided that 
the condition $j\ne i$ is observed. This is straightforward to take
into account in the real-space sum by directly omitting the
terms $i=j$. The Fourier sum, however, expresses the long-range
component as a field which, to be correct at all locations, must contain
the contribution of particle $i$. Measuring the potential at $\vecxi$
using equation~(\ref{eq:ewpecsplit}) will, therefore, include the
self-energy of 
particle $i$ arising from the interaction potential $\varphi_2$. The
self-energy contribution 
from particle $i$ to the Fourier sum is $Gm_i^2\varphi_2(R=0)/2$.
Using equation~(\ref{eq:ewfin}) we obtain, after a little
manipulation, the result 
\begin{eqnarray}
& &U={G\over 2}\Bigg\{\sum_{\scriptstyle i,\jn\atop\scriptstyle \jn\ne i} 
                m_i m_j \varphi_1(\vecxi-\vecxjn) 
   - {\left(\sum_j m_j\right)^2\over
L^3}\,\widetilde\varphi_1(\veck=0) \nonumber \\
& &\hskip 0.9 in+{1\over  L^3} \sum_{i,j,\veck\ne0} 
          e^{i\veck.(\vecxi-\vecxj)} m_i m_j\widetilde\varphi_2(k)
   - \sum_j m_j^2 \varphi_2(R=0)\Bigg\}.
\end{eqnarray}
 
\subsection{Practical issues in computing the Ewald sums}
 
A number of choices are possible for the interaction splitting in
equation~(\ref{eq:ewsplit}). Of primary interest is good convergence
of both the  
real-space and Fourier-space sums. It is also important to choose
splitting functions which allow an efficient numerical scheme to be
constructed. The most commonly used is that given by Ewald (1921), 
but a
number of others have been discussed in the literature (e.g., Nijboer
\& de Wette, 1957).  
 
The Ewald scheme is a one-parameter family with $\varphi_1(R)=-{\rm
erfc}(\eta R)/R$, and $\varphi_2(R)=-{\rm erf}(\eta R)/R$. The Fourier
transform (in three-dimensions) of $\varphi_2$ is
$\widetilde\varphi_2(k)=-4\pi 
e^{-k^2/(4\eta^2)}/k^2$. The parameter $\eta$ is chosen to optimize
convergence. Nijboer \& de Wette~(1957) show that both the real-space
and Fourier-space series have the same rate of convergence if
$\eta=\surd{\pi}/L$. Satisfactory results are obtained for a range of
values near this
result. Machine accuracy (32~bit) can typically be achieved with this
value of $\eta$ by extending the sums in real space and Fourier space
to a radial distance of roughly five image cells or Fourier modes
respectively. The analytic result is independent of the value of
$\eta$ although the rates of convergence of the sums are affected. Increasing
$\eta$ causes a faster convergence of the real-space sum whilst
requiring the accumulation of a greater number of terms in the Fourier
series to achieve a given precision. If $\eta$ deviates
too much from the optimal value, one of the sums will require the
accumulation of a large number of terms for good convergence and we
will be faced with the original computational problem which the
splitting was introduced to solve. 
 
Clearly each computation of the potential (or force etc.) requires
O($N$) operations, and so to find the force on each of $N$ particles,
for example, requires O($N^2$) operations, and so is not competitive
with other methods currently available. The present attraction of the
method is its simplicity and that it enables forces to be calculated
to high precision.

Some improvement in computational efficiency can be obtained by noting
that the Fourier sums in equations~(\ref{eq:ewfin})
and~(\ref{eq:ewtotpot}) can be factorized. 
For equation~(\ref{eq:ewfin}), for example, we can express
$\sum_{j,\veck\ne0} 
e^{i\veck.(\vecx-\vecxj)} m_j \widetilde\varphi_2(k)$ as
$\sum_{\veck\ne0}e^{i\veck.\vecx}\widetilde\varphi_2(k)\,\left(\sum_j
m_je^{-i\veck.\vecxj}\right)$. The sum over $j$ can be precomputed and
stored for the small number of wavenumbers $k$ required (typically a
few hundred). This reduces the cost of the Fourier sum to an O($N$)
operation for a fixed number of wavemodes and allows the parameter
$\eta$ to be increased, which reduces the work of the real-space sum.
Note that since $\varphi_2$ is a real, even function, its Fourier
transform will also be real and even. This allows the complex
exponential to be reduced to a cosine for numerical computation of the
sum. (The factorization just described is then only marginally more
complicated---see section~\ref{subsec:ewform}.)
 
\subsection{Relationship to the P$^3$M method}
 
The P$^3$M algorithm uses an interaction splitting which can be
described using the same terminology set out above.  The splitting
employed is analogous in many ways to choosing a very large value of
$\eta$. This reduces the range over which the real-space sum must be
accumulated and throws most of the work into the Fourier sum. In
P$^3$M the real-space sum is reduced to such an extent that the
effective range is much smaller than the periodic distance,
$L$. Indeed a different functional split is commonly used in which
$\varphi_1$ is compact. This allows efficient techniques to be used in
which only nearby particles need be included in the real-space
sum. Since the effective range of the real-space sum is now much less
than the period distance $L$, this part of the calculation now takes
O($Nn$) operations to compute, where $n$ is the mean number of
particles within the range of the function $\varphi_1$. Provided $n$
does not become too large (as it unfortunately will in gravitational
simulations as clustering develops) the real-space work is essentially
O($N$). 
Pushing much of the work into the Fourier domain is only
advantageous if efficient methods are available for accumulating the
Fourier sum and if we can avoid the convergence problems discussed
above in accumulating forces from the Fourier components. 
 
The Fourier part of the P$^3$M method is best understood in terms of
equation~(\ref{eq:ewpecpot}) rather than equation~(\ref{eq:ewfin}). The
key to the method lies in 
approximating the Fourier components $\widetilde\Phi_{2\,
\veck}$. Instead of calculating afresh the Fourier sum for each value
of $\vecx$ for which it is required, the result is interpolated from
stored values discretized on a regular grid. Provided the field is
adequately sampled, the error in this procedure can be reduced to
acceptable levels. The interaction splitting must be chosen such that
the number of grid points available is sufficient to represent the
harmonic content of the field. Since the number of wavenumbers over
which the field must be known may now be very large (a consequence of the
short range of the real-space sum), an efficient method for calculating
the Fourier components from the density field and interaction
potential must be used. This is achieved by sampling the density
field with a uniform grid and using a Fast Fourier Transform (FFT)
technique for obtaining the potential. Using an FFT also ensures
well-determined convergence properties for the Fourier sums.
 
\subsection{Formul\ae\ for the Ewald method\label{subsec:ewform}}
\eqnote{subsec:ewform}
 
Using the splitting described by Ewald we will now write down explicit 
expressions for equations~(\ref{eq:ewfin}) and~(\ref{eq:ewtotpot})
which can be used to compute  
the potential, its derivatives and the total potential.
 
We have $\varphi_1=-{\rm erfc}(\eta R)/R$, $\varphi_2=-{\rm erf}(\eta
R)/R$ and $\widetilde\varphi_2=-4\pi e^{-k^2/(4\eta^2)}/k^2$. 
(The error function,
${\rm erf}(x)$, is $(2/\surd\pi)\int_0^x e^{-t^2}\,dt$. 
Many approximations to erfc suitable for efficient numerical computation
exist in the literature and are often also available in mathematical
libraries on many current computers.) To calculate
$\widetilde\varphi_1(0)$ we must take the limit of the transform of
$-1/R+{\rm erfc}(\eta R)/R$ as $k\rightarrow0$. This gives
$\widetilde\varphi_1(0)=-\pi/\eta^2$. The value of
$\varphi_2(0)=-\lim_{R\rightarrow0} {\rm erf}(\eta R)/R =
-2\eta/\surd\pi$. Recall that the Fourier modes are labelled by
$\veck=2\pi\vecl/L$ where $\vecl$ is an integer triple. Finally, define 
$C(\vecl)=\sum_j m_j \cos(2\pi\vecl.\vecxj/L)$ and $S(\vecl)=\sum_j 
m_j \sin(2\pi\vecl.\vecxj/L).$ Putting 
all of this together we find:
\begin{eqnarray}
& &\phi(\vecx)=-G\Bigg\{\sum_{\jn} m_j {\rm erfc}(\eta r)/r
\big\vert_{\vec{r}=\vecx-\vecxjn}-{\pi\sum_j m_j\over\eta^2 L^3}\nonumber\\
& &\hskip 0.7in+{1\over \pi L}\sum_{\vecl\ne0} {1 \over l^2} 
\big[C(\vecl)\cos(2\pi\vecl.\vecx/L)+S(\vecl)\sin(2\pi\vecl.\vecx/L)\big]
e^{-\pi^2l^2/(L^2\eta^2)}\Bigg\},\label{eq:ewpotform}
\end{eqnarray}
\eqnote{eq:ewpotform}
and
\begin{eqnarray}
& &U=-{G\over2}\Bigg\{\sum_{i,\jn\atop \jn\ne i} m_i m_j
{\rm erfc}(\eta r)/r\big\vert_{\vec{r}=\vecxi-\vecxjn} - 
{\pi\big(\sum_j m_j\big)^2\over\eta^2 L^3}\nonumber \\
& &\hskip 1.in+{1\over \pi L} \sum_{\vecl\ne0} {1\over l^2}\big[ C^2(\vecl) + 
S^2(\vecl)\big]
e^{-\pi^2l^2/(L^2\eta^2)} - {2\eta\over \surd\pi} \sum_j m_j^2\Bigg\}.
\end{eqnarray}
 
Derivatives of $\phi$ may be calculated trivially from
equation~(\ref{eq:ewpotform}). We have:
\begin{eqnarray}
{\partial\phi\over\partial x_\mu}&=&G\Bigg\{
   \sum_\jn  m_j \left[{\rm erfc}(\eta r) +{2\over\surd\pi}\eta
r e^{-\eta^2r^2}\right]{r_\mu\over
r^3}\Bigg\vert_{\vec{r}=\vecx-\vecxjn}\nonumber\\
& &+{2\over L^2}\sum_{\vec{l}\ne0} {l_\mu\over l^2}
\big[C(\vecl)\sin(2\pi\vecl.\vecx/L)-S(\vecl)\cos(2\pi\vecl.\vecx/L)\big]
e^{-\pi^2l^2/(L^2\eta^2)}\Bigg\},
\end{eqnarray}
and
\begin{eqnarray}
{\partial^2\phi\over\partial x_\mu \partial x_\nu}=& & \nonumber \\
& &\hskip -1 inG\Bigg\{\sum_\jn m_j\left[
-{4\over\surd\pi}\eta^3 e^{-\eta^2r^2} {r_\mu r_\nu\over r^2} +
\left({\rm erfc}(\eta r) + {2\over\surd\pi}\eta r
e^{-\eta^2r^2}\right)
\left({\delta_{\mu\nu}\over r^3}-3{r_\mu r_\nu\over r^5}\right)
\right]_{\vec{r}=\vecx-\vecxjn}\nonumber\\
\noalign{\medskip}
&+&{4\pi\over L^3}\sum_{\vec{l}\ne0} {l_\mu l_\nu\over l^2}
\big[C(\vecl)\cos(2\pi\vecl.\vecx/L) +
S(\vecl)\sin(2\pi\vecl.\vecx/L)\big] 
e^{-\pi^2l^2/(L^2\eta^2)}\Bigg\}.
\end{eqnarray}
 
\noindent Note: asymptotically ${\rm erfc}(x) \sim 1/(\surd\pi x)
e^{-x^2}$.

\end{document}